# Exploiting the Extended π-System of Perylene Bisimide for Label-free Single-Molecule Sensing

Qusiy Al-Galiby, Iain Grace, Hatef Sadeghi and Colin J. Lambert*

Department of Physics, Lancaster University, Lancaster LA1 4YB, United Kingdom

Corresponding author: c.lambert@lancaster.ac.uk

## Abstract

We demonstrate the potential of perylene bisimide (PBI) for label-free sensing of organic molecules by investigating the change in electronic properties of five symmetric and asymmetric PBI derivatives, which share a common backbone, but are functionalised with various bay-area substituents. Density functional theory was combined with a Greens function scattering approach to compute the electrical conductance of each molecule attached to two gold electrodes by pyridyl anchor groups. We studied the change in their conductance in response to the binding of three analytes, namely TNT, BEDT-TTF and TCNE, and found that the five different responses provided a unique fingerprint for the discriminating sensing of each analyte. This ability to sense and discriminate was a direct consequence of the extended π system of the PBI backbone, which strongly binds the analytes, combined with the different charge distribution of the five PBI derivatives, which leads to a unique electrical response to analyte binding.

## Introduction

Partly in response to predictions that silicon technology might soon reach the limits of its evolution [1, 2], there is increasing interest in controlling charge transport across nanometer-scale metal-molecule-metal junctions [3-7] and exploiting the properties of single-molecule devices. Perylene bisimide (PBIs) (also called perylene diimides) [8-11] have emerged as one of the most investigated families of organic molecules, both for their fundamental electronic properties and their industrial applications as dyes and pigments [3, 4]. Their unique properties are primarily derived from their large extended π-systems, which in π-stacked arrays lead to a variety of intermolecular π-orbital overlaps for the different derivatives [12]. Furthermore their high electron affinities, large electron mobilities, their chemical and thermal stabilities, and variety of functional forms with different bay-area substituents has led to their widespread use in organic solar cells [13-16], organic field-effect transistors (OFETs) [17, 18] particularly as n-type materials [19, 20], and organic sensors[21-28].

Recently, it has been experimentally demonstrated that single PBI-based molecules can be attached to gold electrodes and their electrical conductance can bemeasured [29]. In the present paper our aim is to demonstrate that the extended π systems of PBIs make them ideal candidates for the single-molecule, label-free sensing of a variety of analytes. Chemical sensors that work as electronic noses or electronic tongues have attracted extensive attention, because they possess high sensitivity and selectivity towards target analytes, ranging from metal ions and anions to organic neutral chemicals and biological molecules [30-33]. Label-free methods for detecting small molecules are a desirable target technology, because they avoid the need for chemical modification or separation of the analytes, potentially leading to lower costs. Examples of label-free detection include micelle-based bacterial quorum sensing [23] aptamer-based sensing platforms [21], label-free, sequence specific DNA sensing based on fluorescence resonant energy transfer (FRET) [22], nuclear magnetic resonance [24,] nanoplasmonics [25] and surface enhanced Raman spectroscopy (SERS) [26]. However all of these reporting strategies require expensive detectors and are not shrinkable to sub-micron-scale devices and therefore the cost-lowering advantages of label-free sensing are not yet fully realised.

In the present paper, our aim is to demonstrate room-temperature, label-free sensing at the single-molecule level by analysing the variation in electrical properties of the five PBI derivatives shown in figure 1, which possess the same PBI core, but with the following bay-area substituents: *pyrrolidinyl* (aPy-PBI, Py-PBI), *tert-butyl-phenoxy* (P-PBI), *thiobutyl* (S-PBI), and *chlorine* (Cl-PBI). Four molecules (Py-PBI, P-PBI, Cl-PBI, S-PBI) possess pyridyl anchor groups at opposite ends and are therefore symmetric. The fifth molecule (aPy-PBI) has a *pyridyl* anchor group on the top and a *cyclohexyl* anchor group on the bottom [29] and is asymmetric.

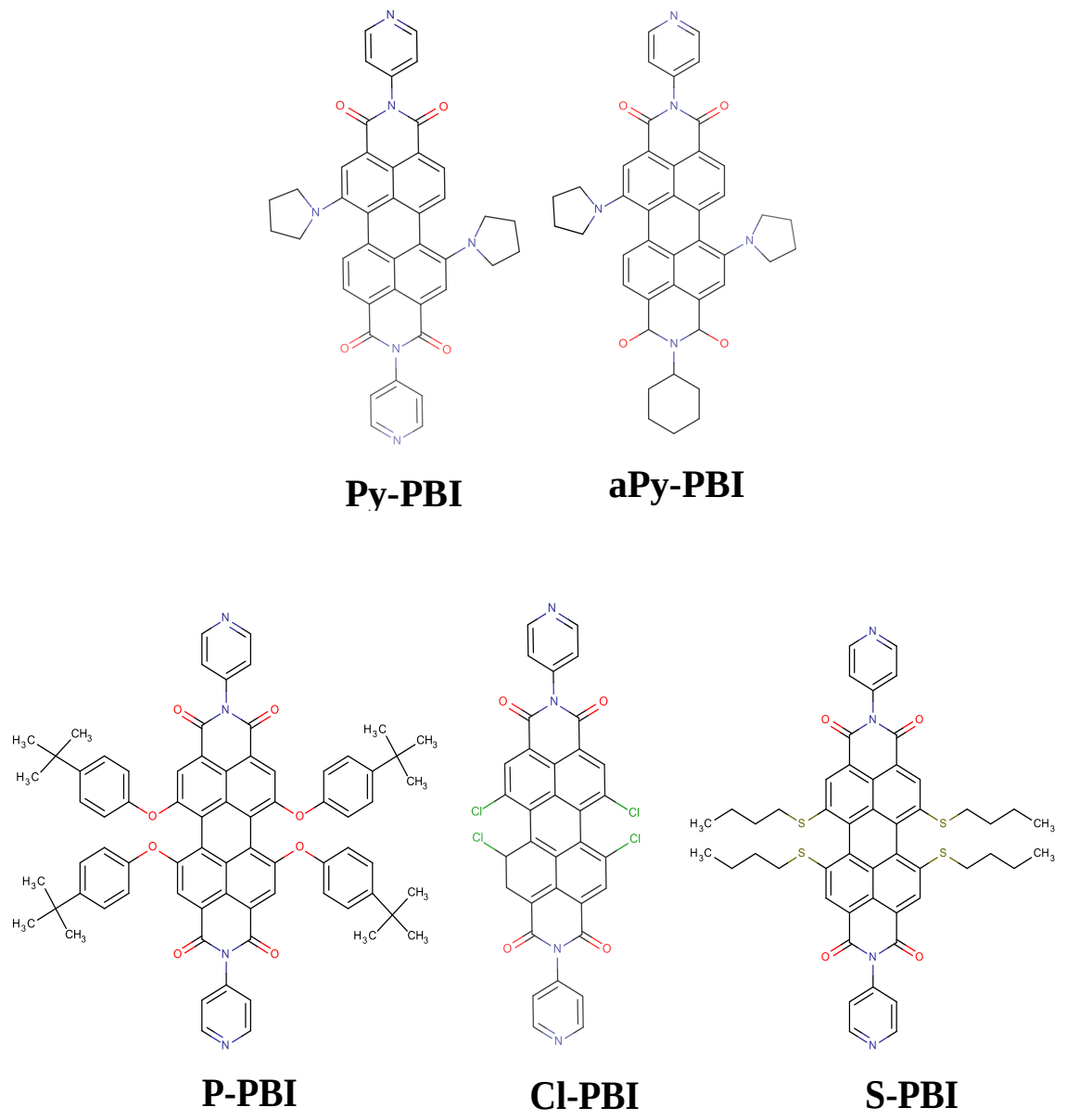

**Figure 1.** PBI-based molecular structures: Py-PBI, (aPy-PBI), P-PBI, Cl-PBI. and S-PBI.

We shall demonstrate that these PBIs can sense and discriminate between the three organic analyte molecules shown in figure 2. The chosen analytes are TNT and TCNE which are well known electron acceptors [34, 35] and the donor BEDT-TTF [36-38], whose $\pi$-donor capability is weaker than that of TTF [39]. Our aim is to demonstrate that the presence of both acceptors and donor molecules alter the conductance of a single molecule of PBI and that the combined responses of the PBIs form a unique fingerprint that can be used to discriminate between the analytes.

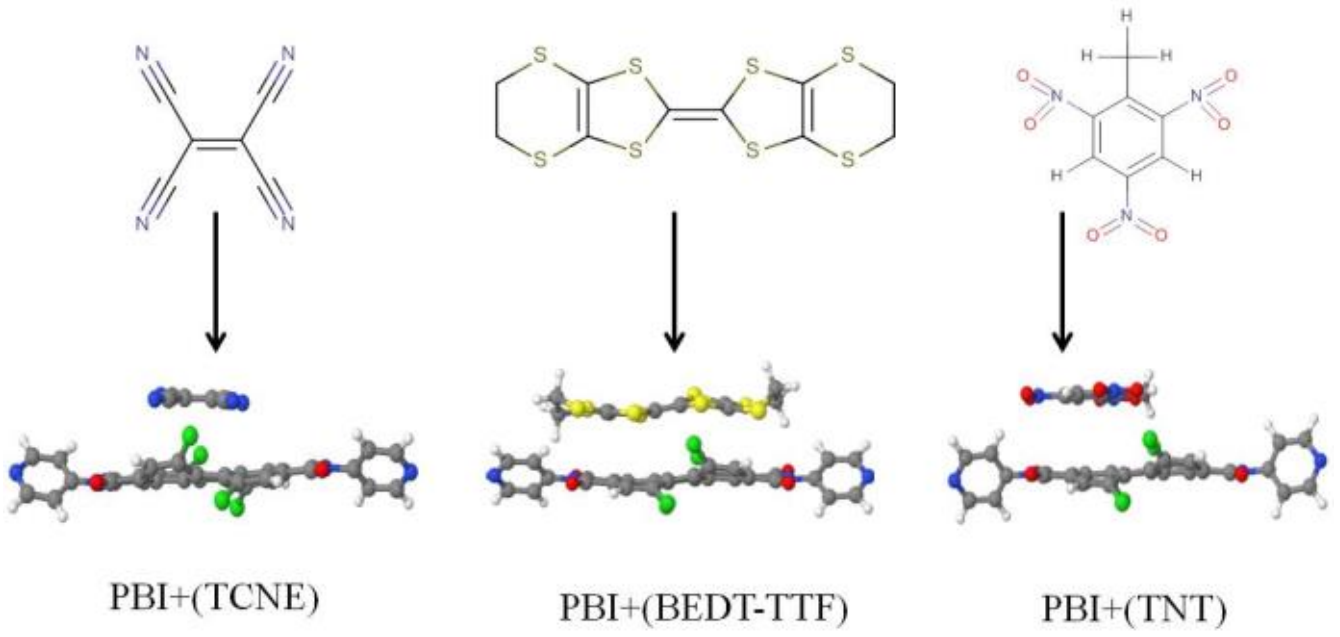

**Figure 2.** The DFT calculated optimum geometries for molecular complexes of TCNE, BEDT-TTF, and TNT with the PBI molecule.

## Computational Methods

To calculate electrical properties of the molecules in figure 1, the relaxed geometry of each molecule was found using the density functional (DFT) code SIESTA [40] which employs Troullier-Martins pseudopotentials to represent the potentials of the atomic cores [41], and a local atomic-orbital basis set. We used a double-zeta polarized basis set for all atoms and the generalized gradient approximation (GGA-PBE) for the exchange and correlation functionals [42, 43]. The Hamiltonian and overlap matrices are calculated on a real-space grid defined by a plane-wave cutoff of 150 Ry. Each molecule was relaxed to the optimum geometry until the forces on the atoms are smaller than 0.02 eV/Å and in case of the isolated molecules, a sufficiently-large unit cell was used, for steric and electrostatic reasons. All PBI molecules with anchor groups and bay-area substituents were found to be twisted after relaxation. Figure S1 of SI shows that the frontier molecular orbitals of all five PBI derivatives are delocalized across their central backbones.

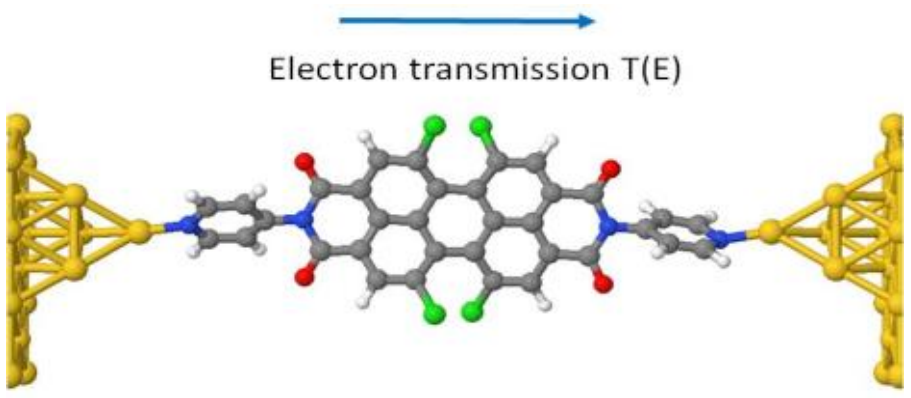

**Figure 3.** An example of an optimized configuration of the system containing a single molecule (Cl-PBI) attached to two metallic leads. Further details for the rest optimized configurations of PBIs with the gold leads are in figure S2 of SI.

After obtaining the relaxed geometry of each isolated molecule, the molecules were then placed between gold electrodes, as shown in figure 3. For structures such as that shown in figure 3, the central region of the junction is composed of a single molecule attached to two gold (111) leads. The equilibrium distance between the N atom of each pyridyl anchor group and the centre of the apex atom of each gold pyramid was initially 1.9 Å. After geometry optimization the distance changed slightly from the initial value to a final value of 2.05 Å. Similarly, the distance between the cyclohexyl group of asymmetric molecule (aPy-PBI) and the centre of the apex atom of gold was found to be 2.89 Å.

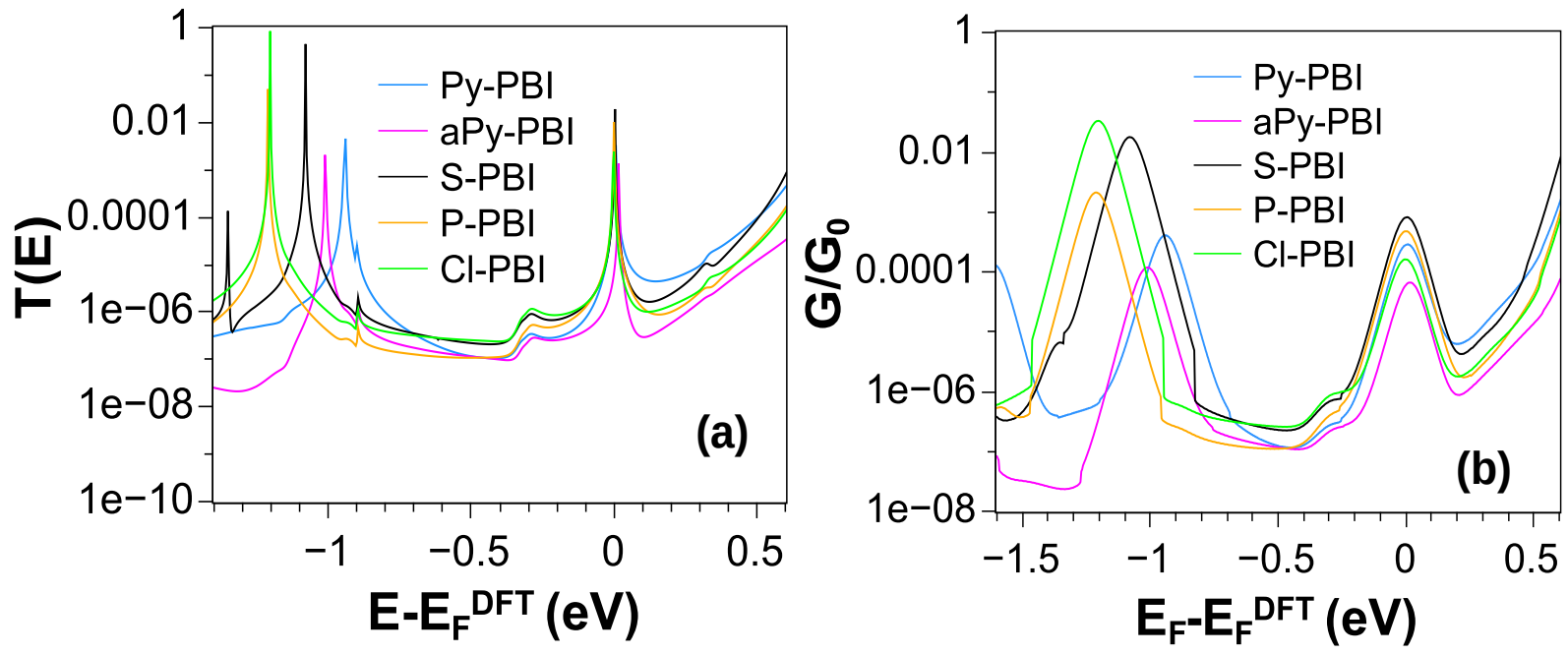

**Figure 4.** (a) Transmission coefficients as a function of energy for the five perylene bisimides. (b) Room-temperature conductance as a function of the Fermi energy.

For each relaxed structure, such as figure 3, we use the GOLLUM method [44] to compute the transmission coefficient *T(E)* for electrons of energy *E* passing from the left gold electrode to the right electrode. GOLLUM is a next-generation code, born out of the non-equilibrium transport code SMEAGOL code [45] and is optimised for computing scattering properties of a range of quantum systems such as molecular junctions. Once *T(E)* is computed, we calculate the zero-bias electrical conductance *G* using the Landauer formula:

$$G = I/V = G_0 \int_{-\infty}^{\infty} dE\, T(E)(-\frac{df(E)}{dE}) \qquad (1)$$

where $G_0 = \left(\frac{2e^2}{h}\right)$ is the quantum of conductance, and $f(E)$ is Fermi distribution function defined as $f(\mathrm{E}) = [e^{(E-E_F)/k_B\mathrm{T}} + 1]^{-1}$ where $k_B$ is Boltzmann constant $k_B = 8.62 \times 10^{-5} eV/k$ and $T$ is the temperature. Since $f(\mathrm{E})$ is a function of the Fermi energy $E_F$, the conductance G is a function of the energetic locations of the molecular orbitals of the PBI-analyte complexes relative to $E_F$. Since the quantity $(-df(E)/dE)$ is a normalised probability distribution of width approximately equal to $k_BT$, the above integral represents a thermal average of the transmission function *T(E)* over an energy window of the width $k_BT$ (= 25meV at room temperature). In what follows, we shall demonstrate that either the zero-bias conductances or the current-voltage (I-V) relations of each complex can be used to discriminate between the different analytes. To compute the electrical current *I* at finite voltage *V* we use the expression [33,46]:

$$I = \left(\frac{2e}{h}\right) \int_{-\infty}^{\infty} dE\, T(E)[f_{left}(\mathrm{E}) - f_{right}(\mathrm{E})] \qquad (2)$$

where $f_{left}(\mathrm{E}) = [e^{(1/K_BT)(E-E_F^{left})} + 1]^{-1}$ and $f_{right}(\mathrm{E}) = [e^{(1/K_BT)(E-E_F^{right})} + 1]^{-1}$, with $E_F^{left} = E_F + \frac{eV}{2}$ and $E_F^{right} = E_F - \frac{eV}{2}$.

## Results and discussion

To benchmark our calculations, we first compared our theoretical predictions with experimental measurements of ref. [29], where it was found that the electrical conductances of the bare PBIs are ordered as follows: Cl-PBI < PBI < S-PBI < aPy-PBI < Py-PBI. For each of these molecules, figure 4a shows the transmission coefficients *T(E)* as a function of energy *E*, relative to the DFT-predicted Fermi energy $E_F^{DFT}$. Figure 4a shows that the effect of the bay substituent atoms is to shift the position of the HOMO resonances of the bare PBIs, while the LUMOs remain pinned close to the Fermi energy. The corresponding room-temperature conductances as a function of $E_F$ are shown in figure 4b.

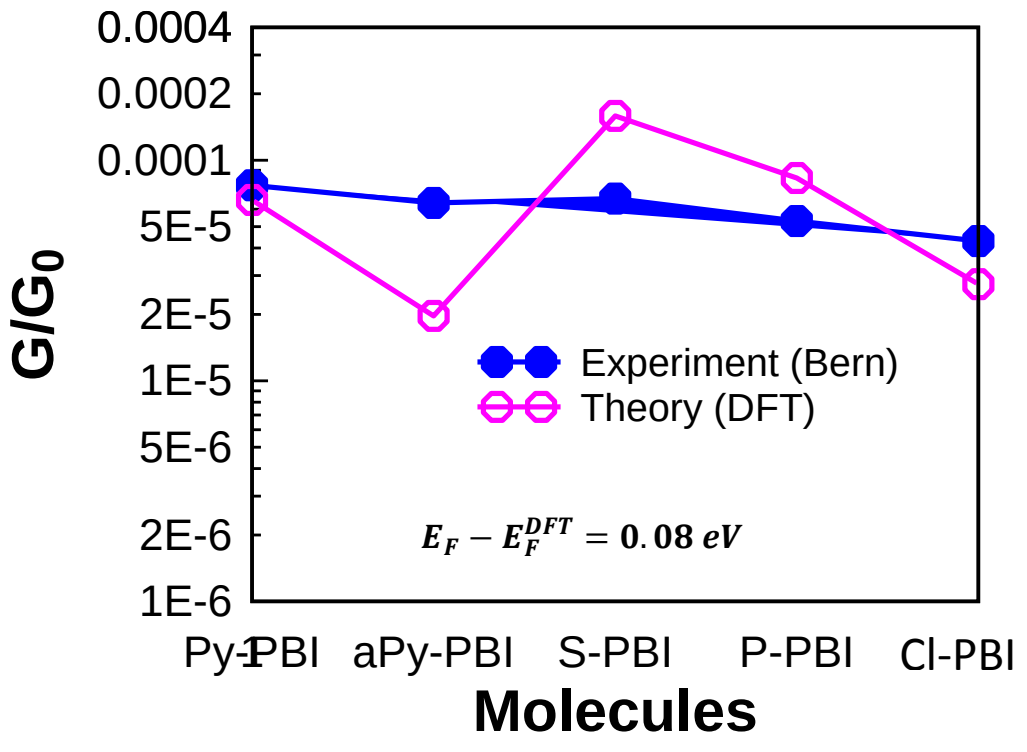

**Figure 5.** Comparison between the experimental [29] and theoretical conductances of the bare PBIs, obtained with $E_F$ shifted by 0.08eV relative to the bare DFT value.

It is well-known that *DFT* can give an incorrect value for the Fermi energy and therefore we use equation (1) to evaluate the room-temperature conductance for a range of values of $E_F$ in the vicinity of $E_F^{DFT}$ and then choose the value of $E_F$ which yields the closest agreement with experiment. For a given value of $E_F$ , the five theoretical conductances obtained from figure 4b were compared with the experimentally-measured conductances [29] and a value of $E_F = E_F^{DFT} + 0.08\,eV$ was chosen to yield the closest fit to experiment. (See SI for more details.) Figure 5 shows the resulting comparison between experiment and theory.

## Molecular Complexes

For the five PBI molecules in figure 1, the formation of molecular complexes with each of the three different analytes TCNE, BEDT-TTF and TNT was investigated using DFT. In each case, after geometry relaxation, the charge transferred between the two molecules and the binding energy was computed (This data is shown in tables S1, S2 and S3 of the SI) The calculations show that TCNE and TNT gain electrons from the backbone of PBI, but the BEDT-TTF donates electrons to the backbone. As an example, figure 6 shows the optimised TCNE-(Py-PBI) complex, in which TCNE pi-stacks a distance Y=0.37 nm above the perylene. The details of all optimised PBI-adsorbate complexes attached to gold electrodes are shown in figures S12, S13, and S14 of the SI.

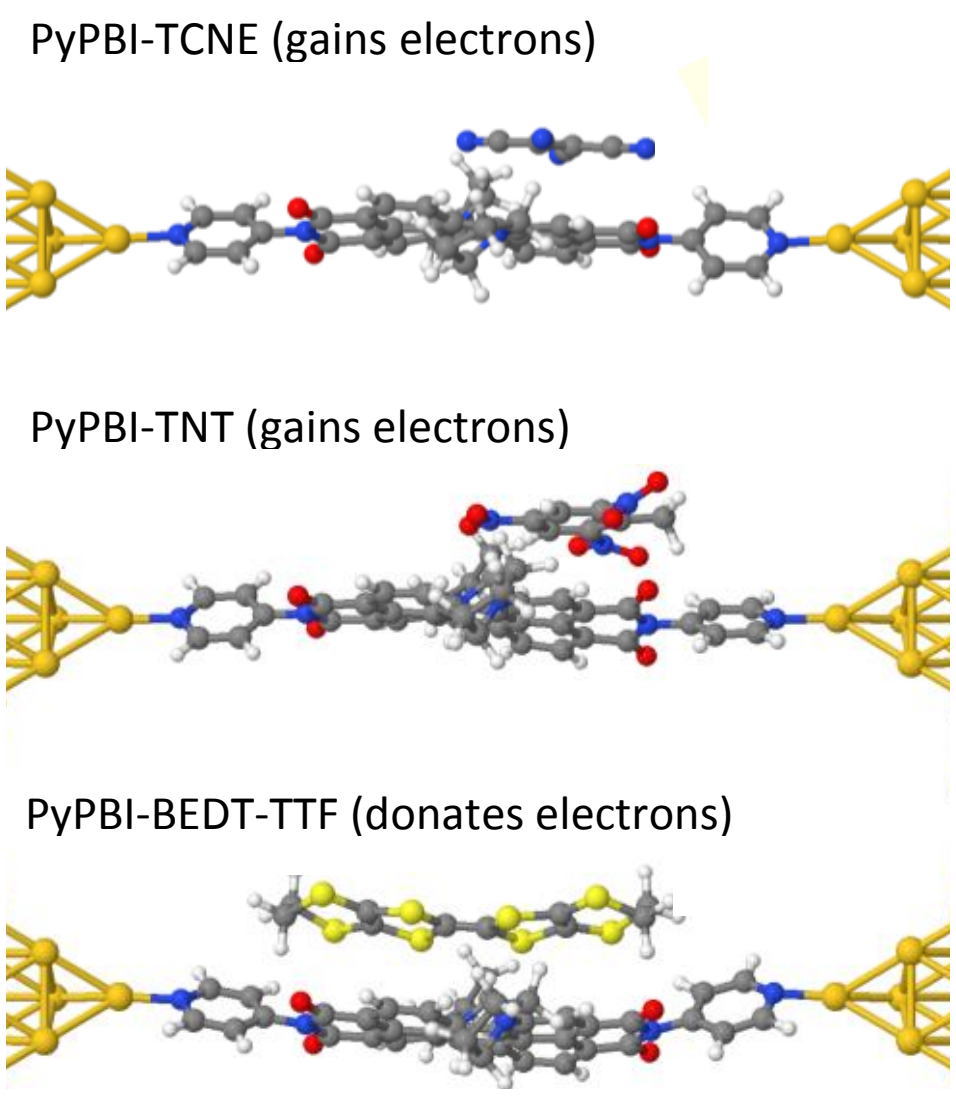

**Figure 6.** Optimized configuration Py-PBI with (TCNE, TNT, and BEDT-TTF) which obtained from optimum configuration attached to two metallic leads.

Figures 7a-c show the transmission coefficients *T(E)* due to the binding of each of the three analyte molecules on Py-PBI. For the two acceptor molecules TCNE and TNT, there is a shifting of the HOMO resonance and extra resonances appear close to the LUMO. For the donor BEDT-TTF a resonance also appears in the HOMO-LUMO gap, accompanied by additional resonances below the HOMO. Figure 7d shows the corresponding I-V characteristics of each of the complexes, demonstrating that for optimally-bound complexes, the distinct I-V curves can be used to discriminate between BEDT-TTF and the other analytes, although the difference between TCNE and TNT is smaller.

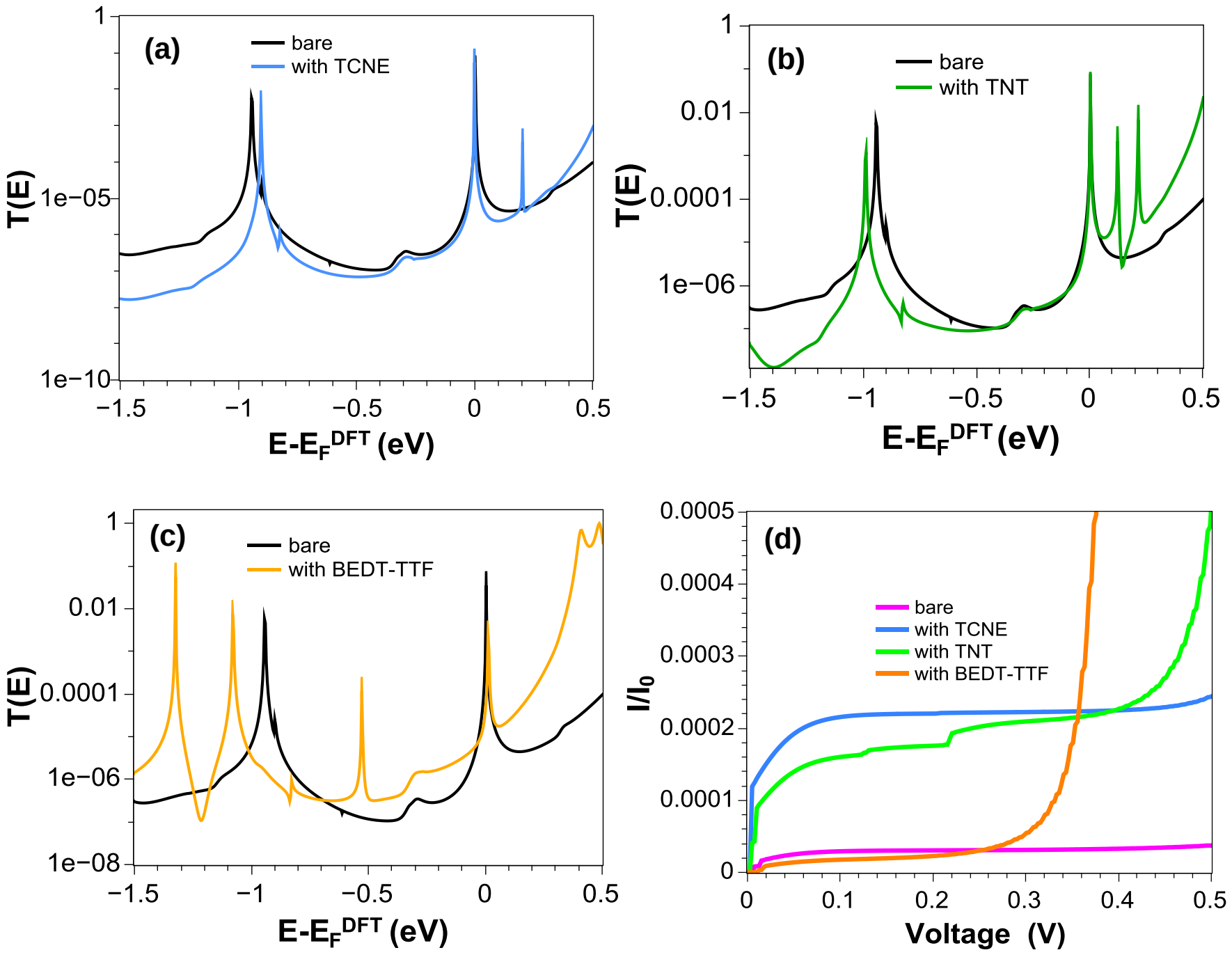

**Figure 7.** Transmission coefficients as a function of energy at T=0K for optimum configuration of Py-PBI with (a) TCNE , (b) TNT, and (c) BEDT-TTF. Figure (d) shows the current as a function of voltage at T=300K for the bare Py-PBI and in the presence of the three analytes (TCNE, TNT, and BEDT-TTF).

The results of figure 7 correspond only to optimally-bound analytes. At room temperature, the adsorbates are subject to thermal fluctuations and will sample many positions across the PBI backbones. To investigate the role of fluctuations, for each PBI molecule we repeated the above calculations for 214 configurations of each adsorbate. The results for the Py-PBI molecule are shown in figures 8a-c. For illustrative purposes, we only show 35 configurations (the results for all 214 transmission curves are shown in Fig S5-S9 of SI). In each case we see that fluctuations in the position of the analyte cause the transmission resonances to shift over a range of energies. In many cases, these possess a characteristic Fano line shape, associated with the interaction between localized states on the analytes and extended states on the PBI backbones [47, 48]. After computing the ensemble average of these curves, the resulting the ensemble-averaged, room-temperature current as a function of voltage is shown in figure 8d, along with the standard deviation in their means (see SI for more details). This demonstrates that even in the presence of fluctuations, BEDT-TTF produces a distinctly different I-V curve from the other two analytes.

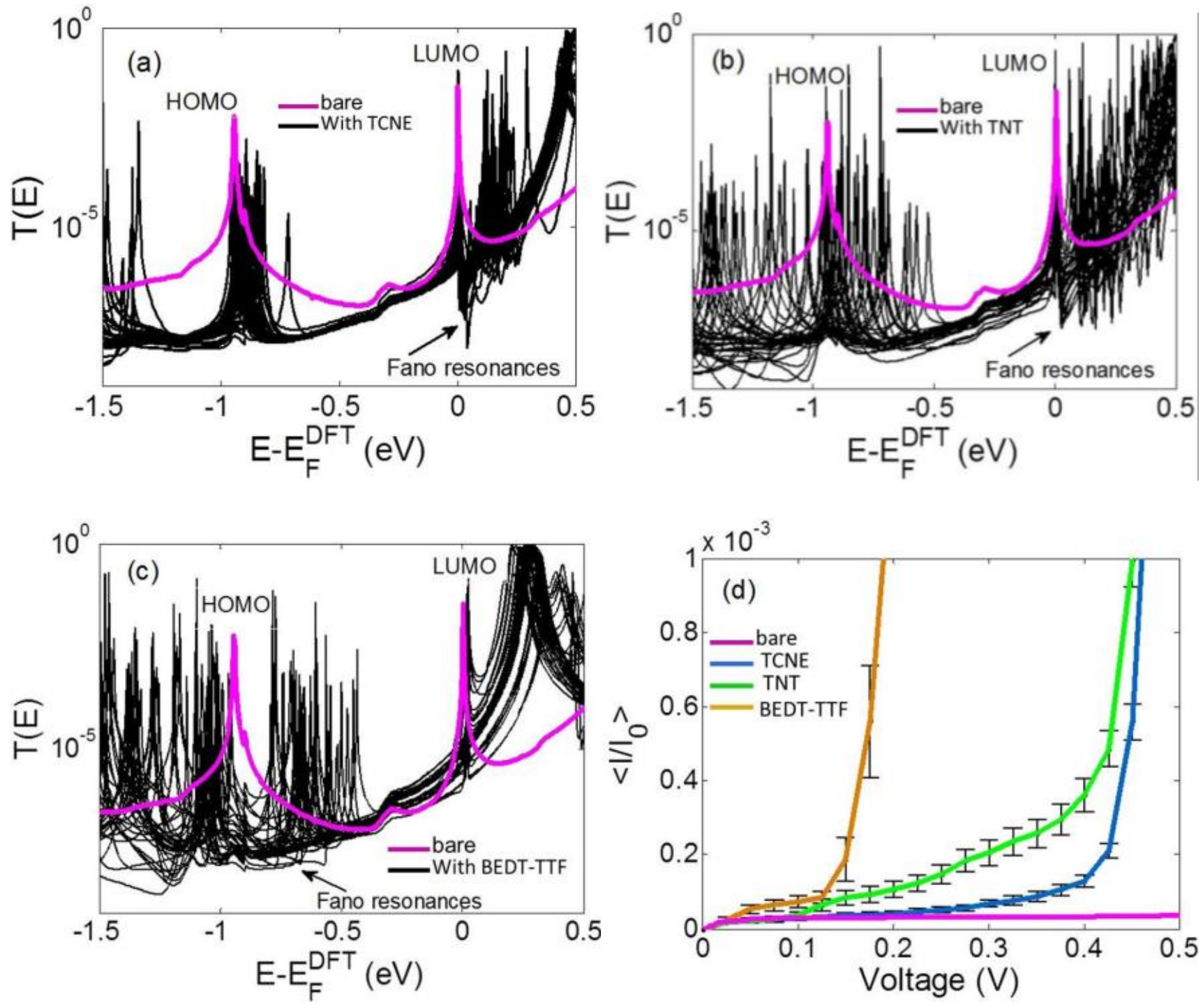

**Figure 8.** Illustrates the transmission curves of 35 fluctuation position in case Py-PBI to explain the effect of adsorption of analytes molecules (a) TCNE, (b) TNT, (c) BEDT-TTF, and (d) shows the ensemble-averaged of current as a function of voltage at T=300K for bare Py-PBI and in the presence of the three analyte molecules (TCNE, TNT, and BEDT-TTF), where the error bars in figure 8d shows the standard deviation in the means of the currents.

The corresponding room-temperature, ensemble-averaged zero-bias conductances for all analytes and backbone molecules are shown in figure 9. This demonstrates that not only I-V curves, but also the conductance of each PBI backbone changes due to the adsorption of a single analyte molecule. For any one PBI derivative (eg P-PBI TNT compared with BEDT-TTF), the change in conductance upon binding may not discriminate between all analytes. However, the spectrum of five conductance changes forms a unique fingerprint, which can be used to discriminate between all three analytes.

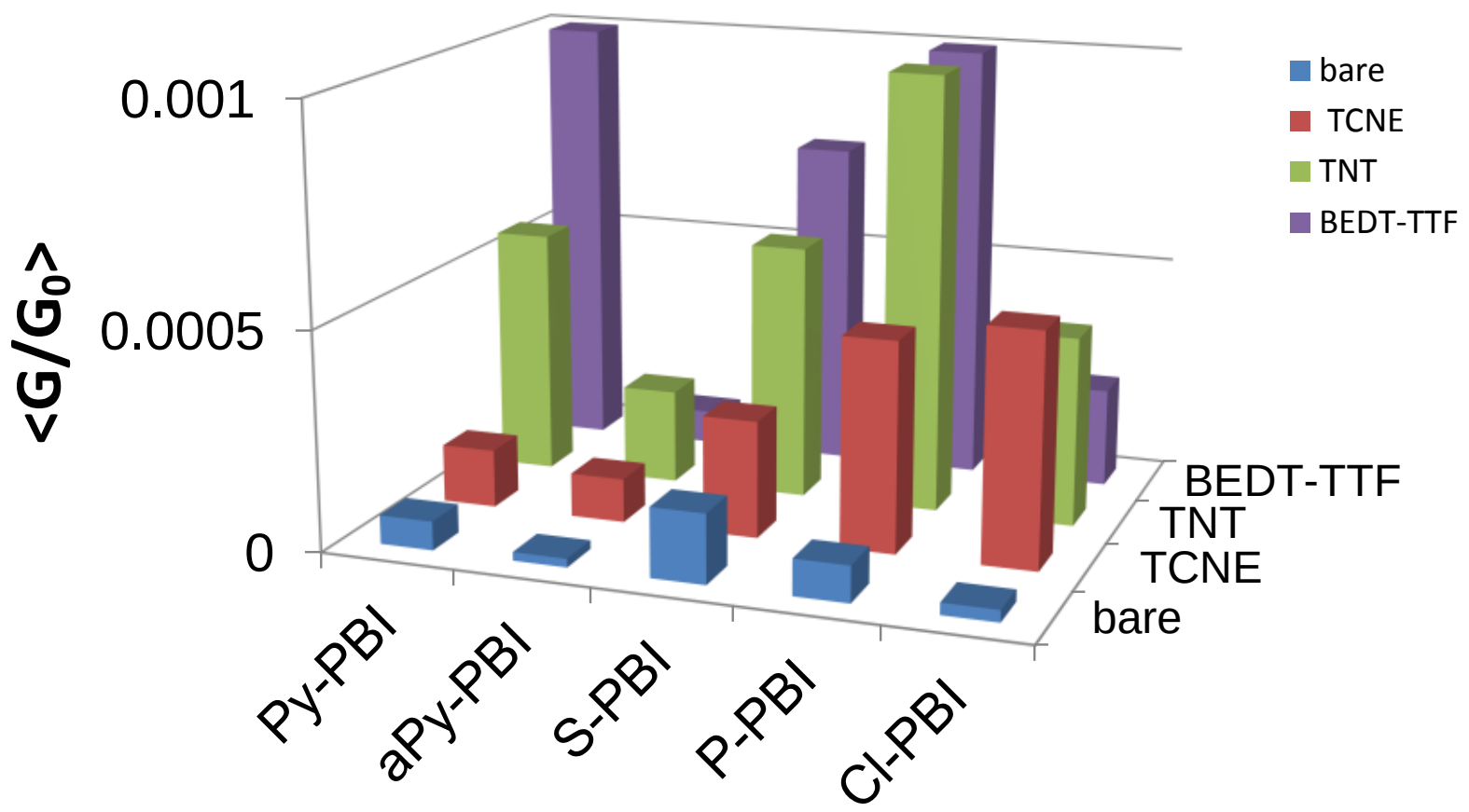

**Figure 9.** The room-temperature, ensemble-averaged conductance across the PBI family due to charge transfer complex formation (obtained from the average of the 214 configurations of the TCNE, BEDT-TTF and TNT) around the backbone of five PBIs.

## Conclusion

We have presented a study of the electron transport properties of five PBI derivatives with various bay-area substituents: pyrrolidinyl (aPy-PBI, Py-PBI), tert-butyl-phenoxy ( P-PBI), thiobutyl (S-PBI), and chlorine (Cl-PBI). Recent experimental papers measuring the electrical conductance of backbone molecules in different solvent have shown that the conductance can be changed by the solvent, even when the solvent does not forma complex with the backbone49. In our case, the analytes bind to the backbones with up to 1 eV of binding energy and the conductance changes are much larger. Our results compare well with recent experimental values for the electrical conductance of these molecules in the absence of adsorbates. We analyzed the change in conductance of these molecules when single molecules of three analytes (TCNE, TNT, and BEDT-TTF) were adsorbed onto the PBI backbones and found that the resulting changes in the ensemble-averaged room-temperature conductances produced a unique fingerprint for each analyte, which allows their discriminating sensing at the single molecule level. These conductance changes arose from a combination of charge redistribution associated with charge-transfer-complex formation, and the formation of a Fano resonances associated with the interaction of a bound state on the adsorbate and extended orbitals of the PBI backbones.

## Acknowledgements

This work is supported by UK EPSRC grants EP/K001507/1, EP/J014753/1, EP/H035818/1, the European Union Marie-Curie Network MOLESCO 606728 and the Iraqi Education Ministry, Al Qadissiya University.

Supplementary Information.

# Exploiting the Extended π-System of Perylene Bisimide for Label-free Single-Molecule Sensing

*Qusiy Al-Galiby, Iain Grace, Hatef Sadeghi and Colin Lambert**

*Physics Department, Lancaster University, LA1 4YB UK*

**c.lambert@lancaster.ac.uk*

Py-PBI

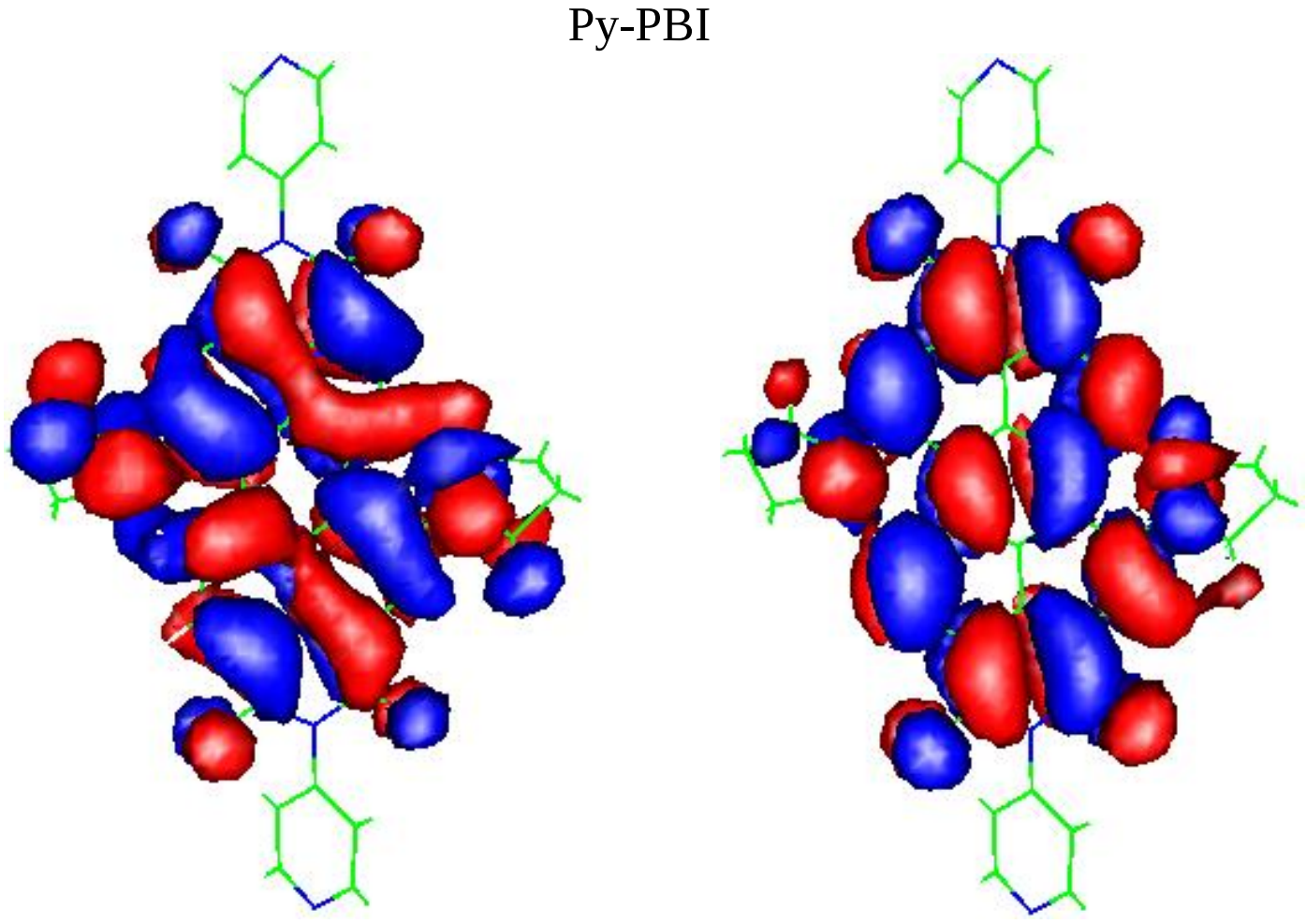

HOMO LUMO

aPy-PBI

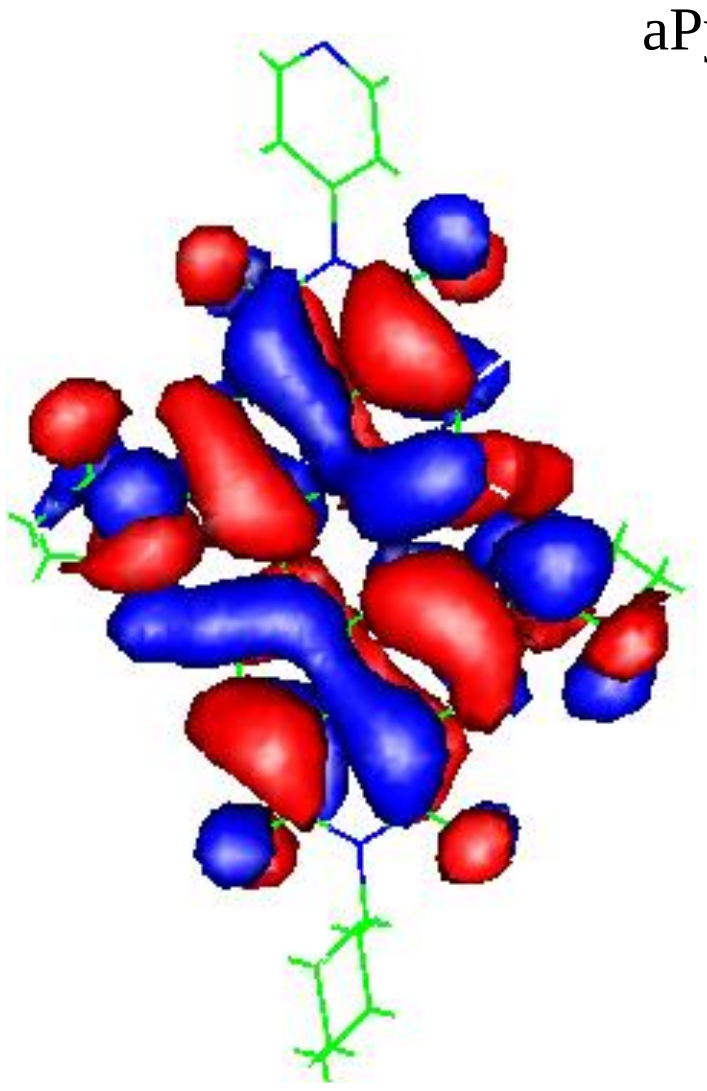

HOMO

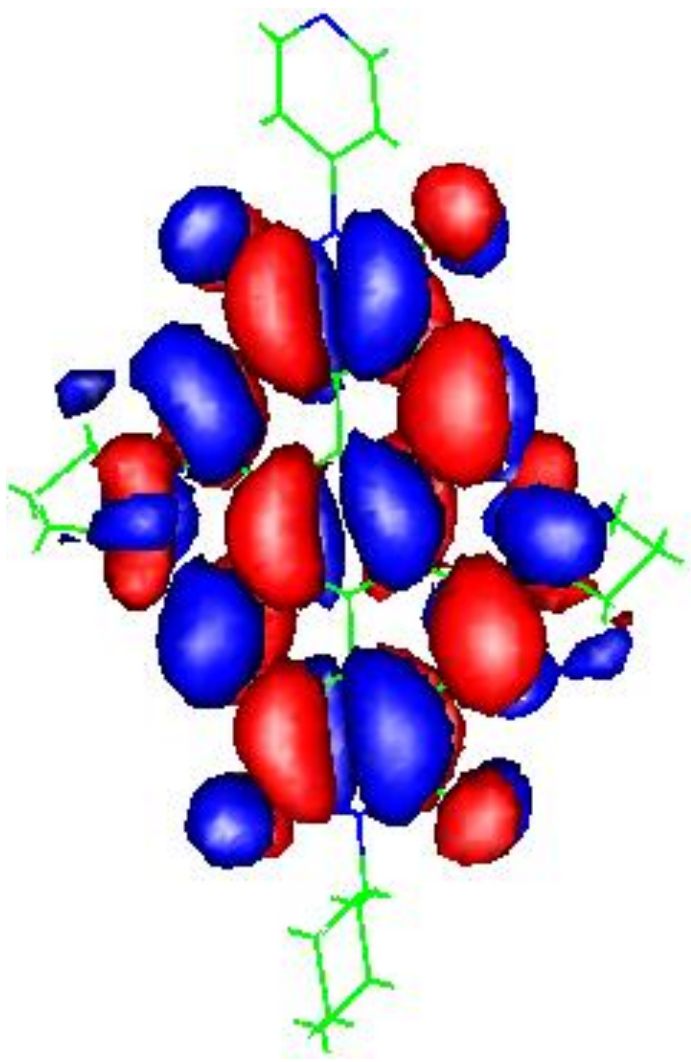

LUMO

S-PBI

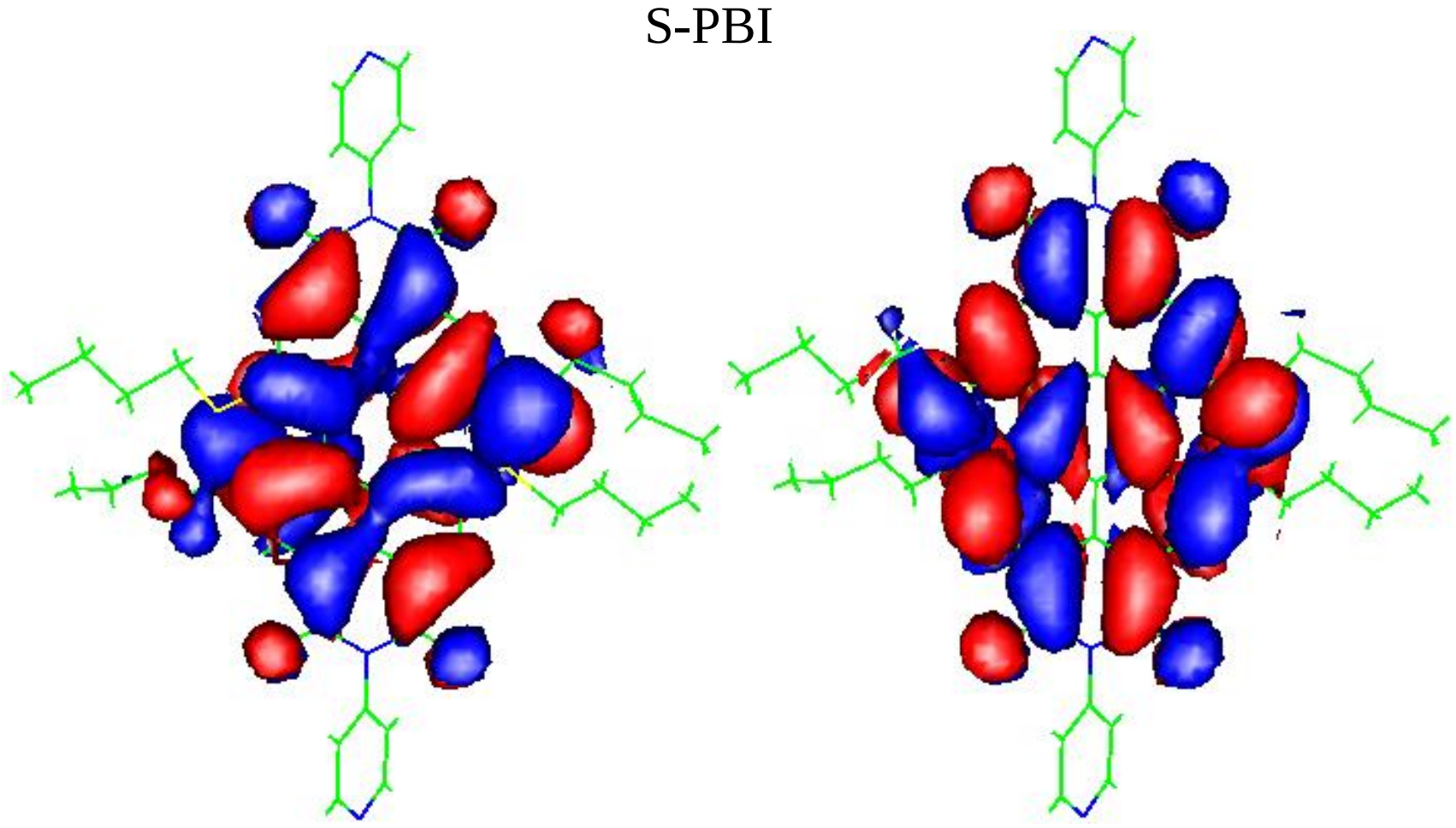

HOMO LUMO

P-PBI

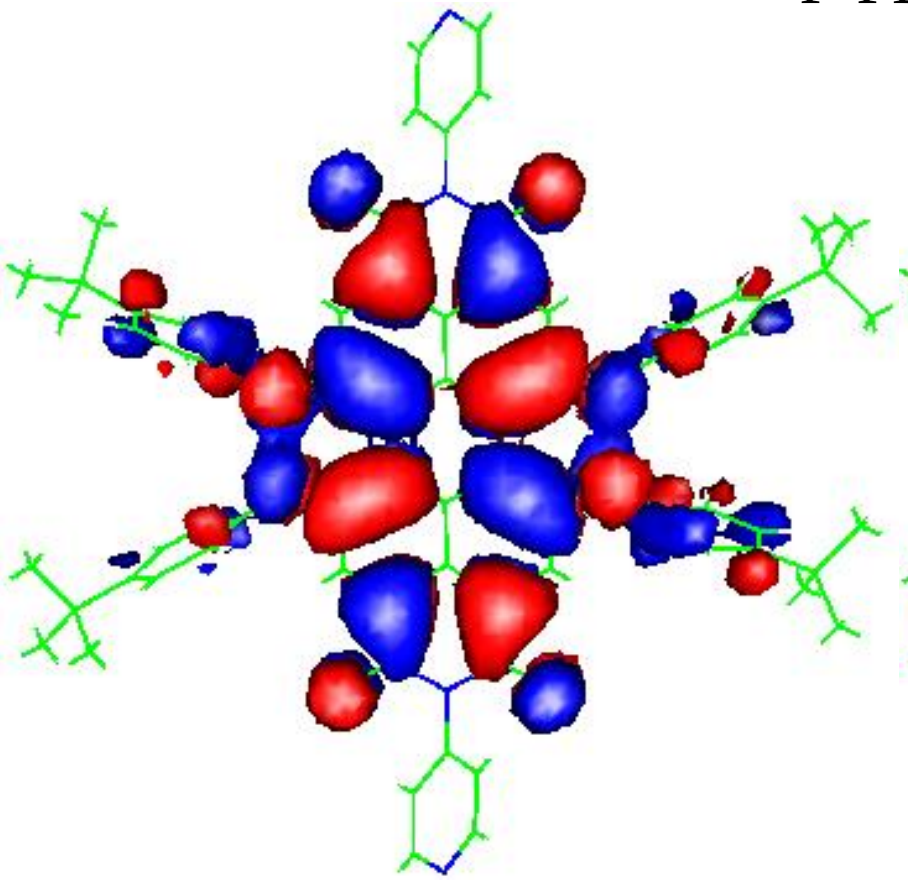

HOMO

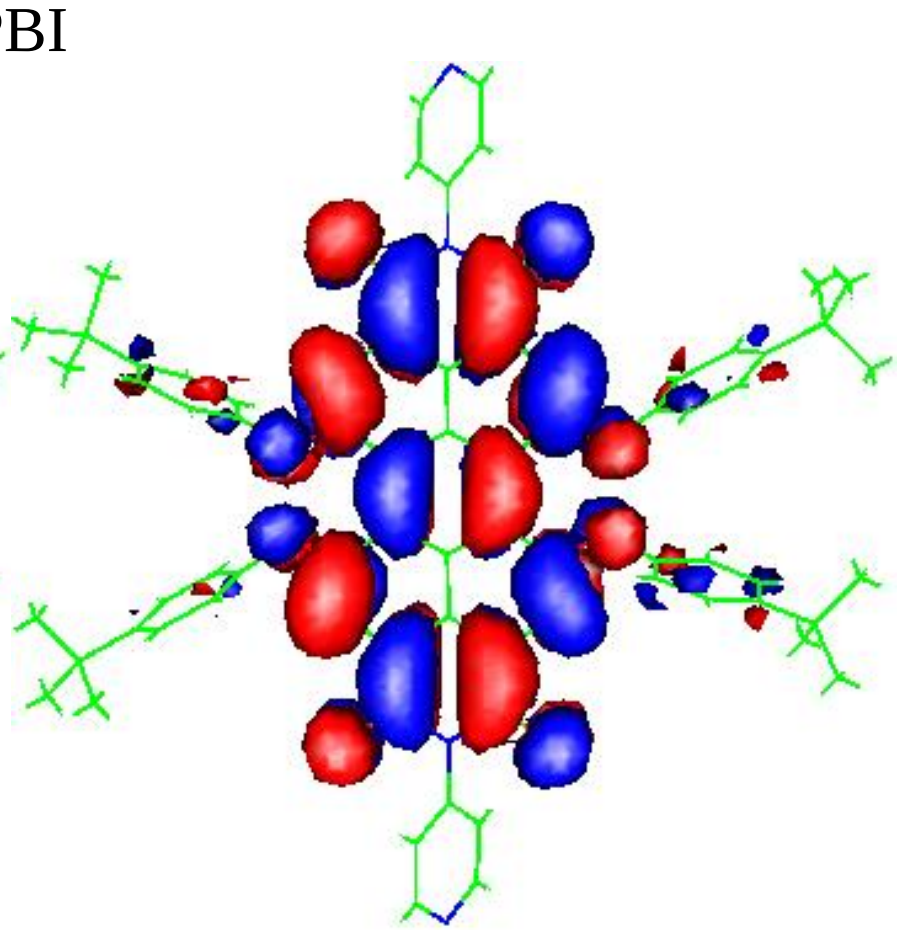

LUMO

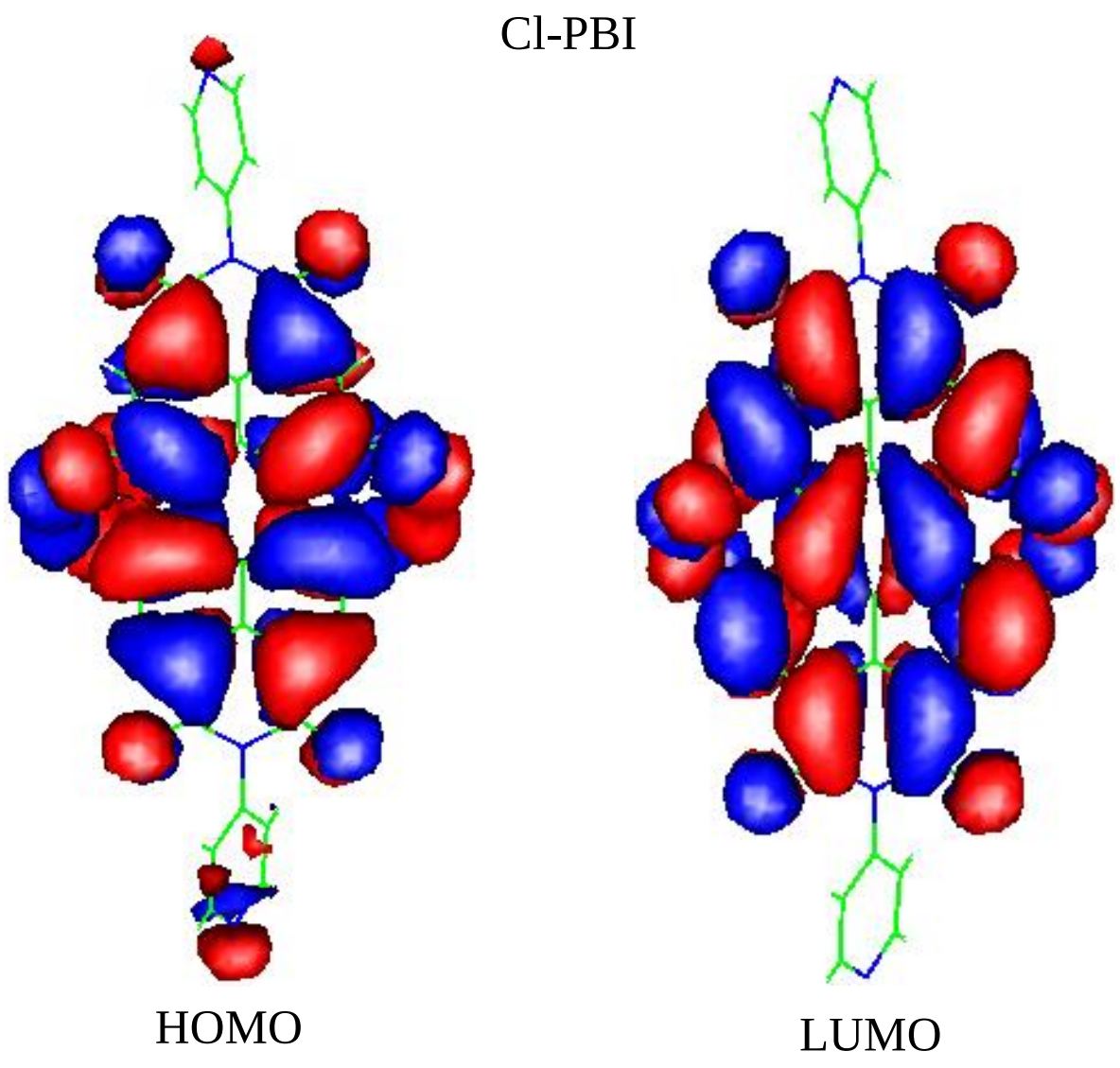

**Figure S1.** Frontier molecular orbitals of the all perylene bisimide derivatives obtained using the DFT code SIESTA. Red corresponds to positive and blue to negative regions of the wave functions.

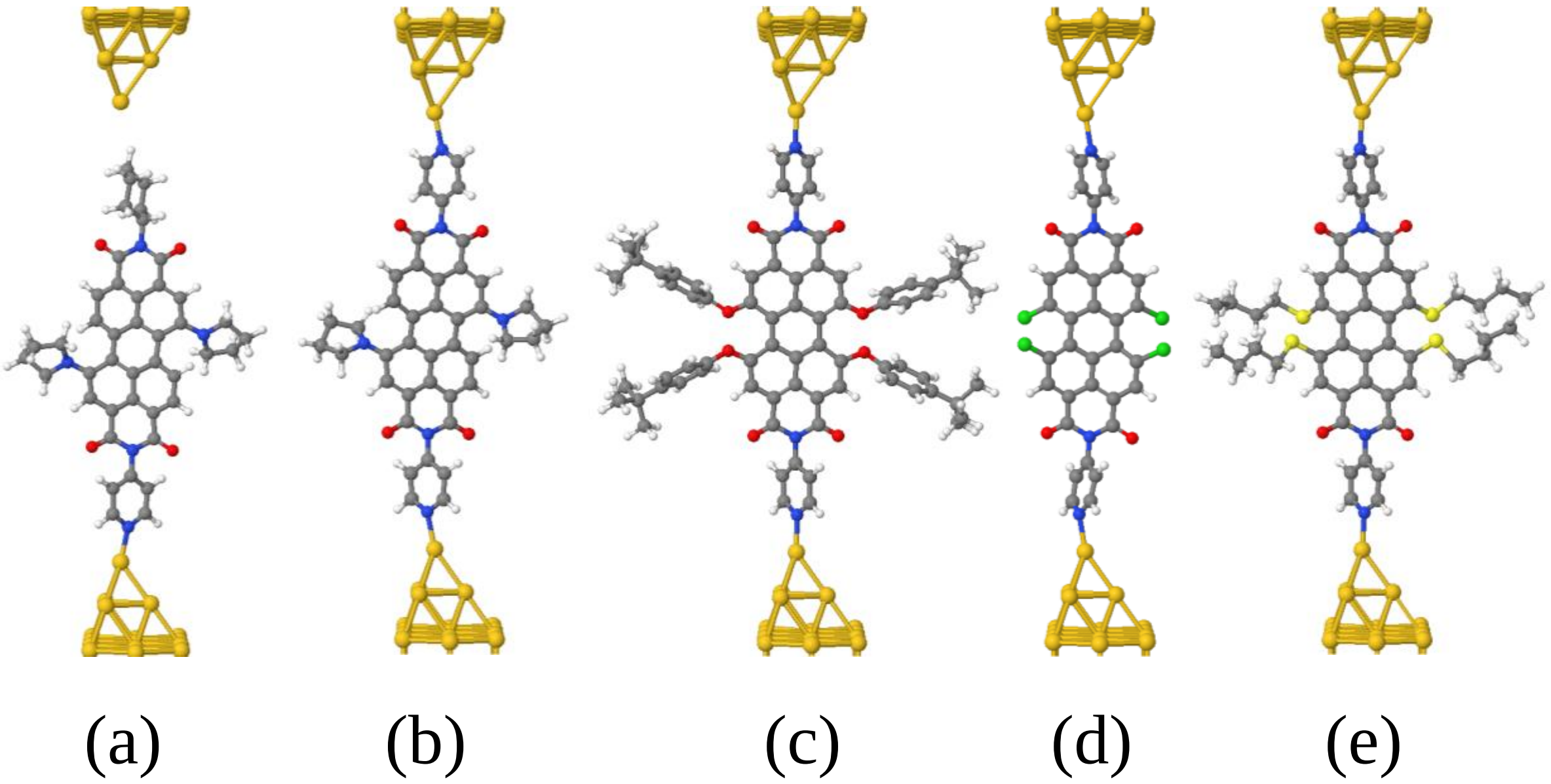

**Figure S2.** Optimized configurations of the single molecules (a) aPy-PBI, (b) Py-PBI, (c) P-PBI, (d) Cl-PBI, and (e) S-PBI attached to two metallic leads.

## The origin of Fano resonaces

To illustrate the origin of Fano resonances, consider the sketch showing a bound state (blue) of energy ε, coupled to an extended backbone state (brown) of energy $\varepsilon_1$ by a coupling matrix element $\alpha$ .

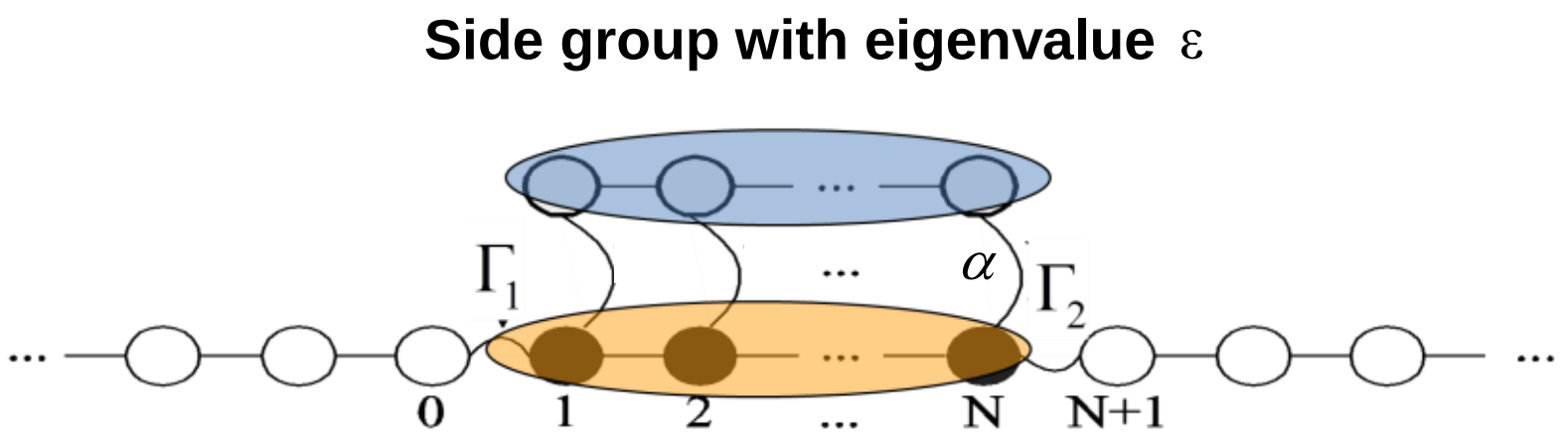

It can be shown [33] that the transmission coefficient for such a combination is given by

$$T(E) = \frac{4\Gamma_1\Gamma_2}{\left(E - \varepsilon_1 - \frac{\alpha\alpha^*}{E-\varepsilon}\right)^2 + (\Gamma_1 + \Gamma_2)^2} \qquad (S1)$$

A plot of this expression is shown in figure S5.

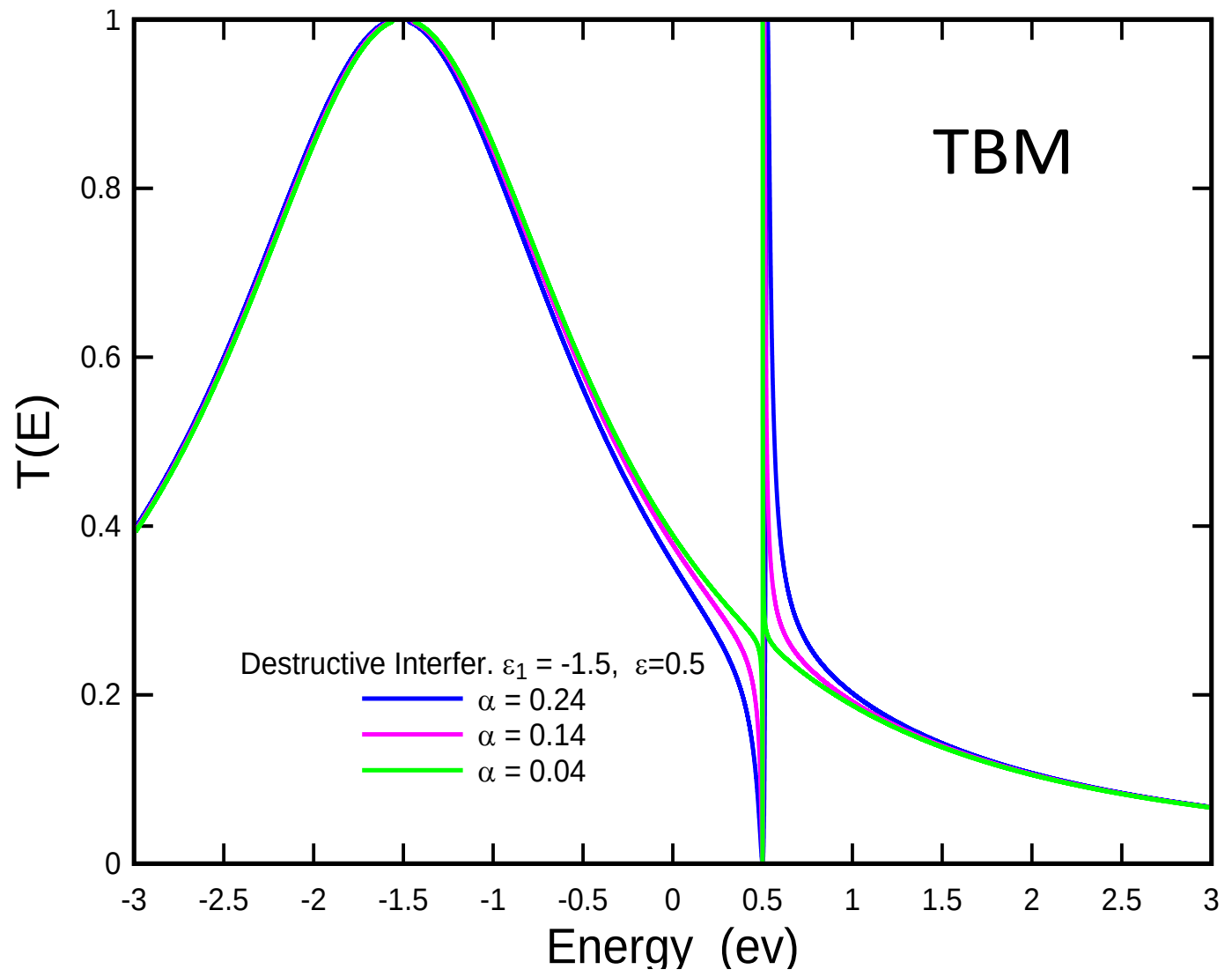

**Figure S3.** A plot of equation S1 for values of the level broadening due to the contacts of $\Gamma_1 = 0.1, \Gamma_2 = 0.1$.

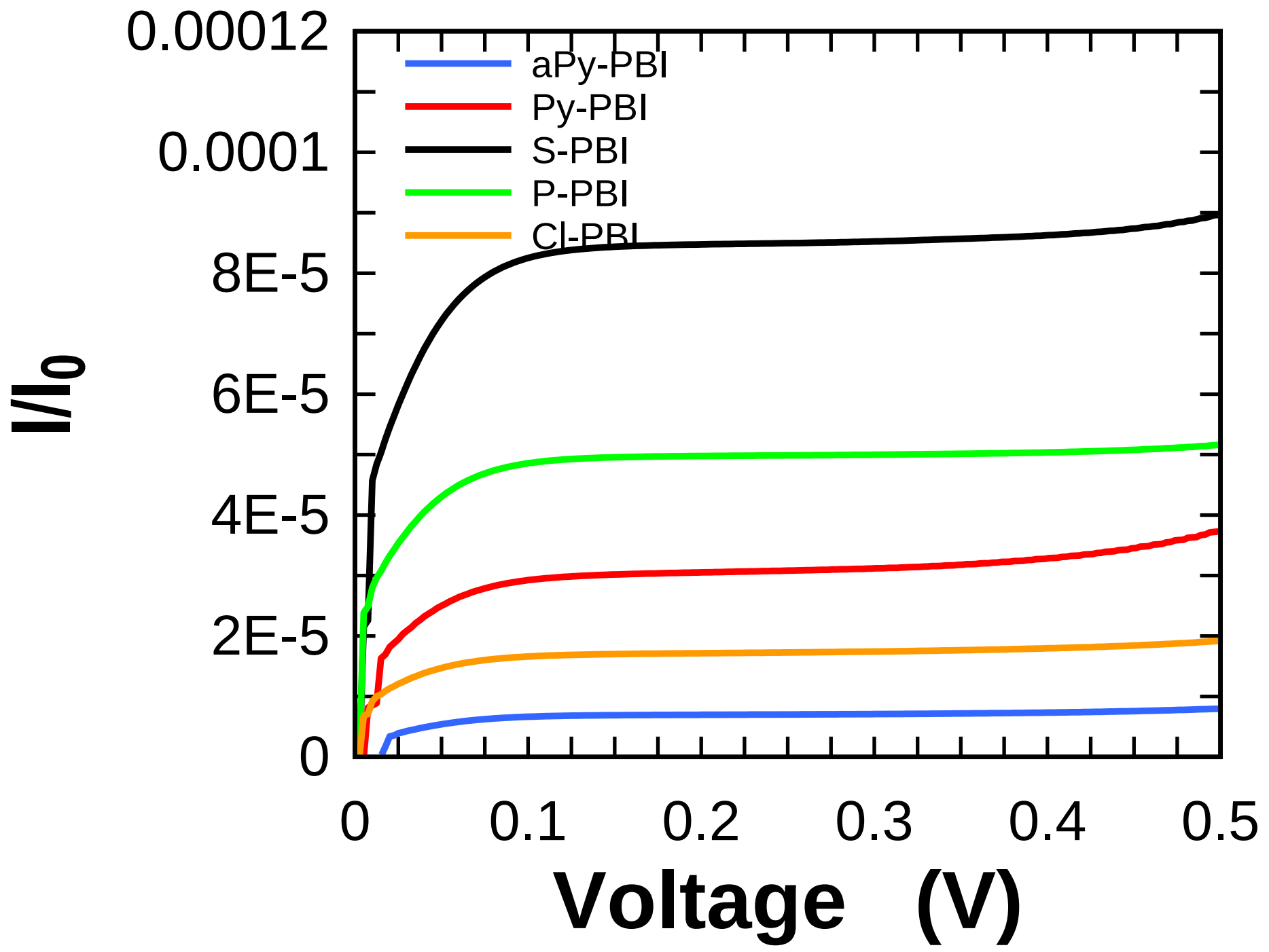

**Figure S4.** DFT calculations for the corresponding results for the room-temperature current as a function of the voltage for the five perylene bisimides in bare case.

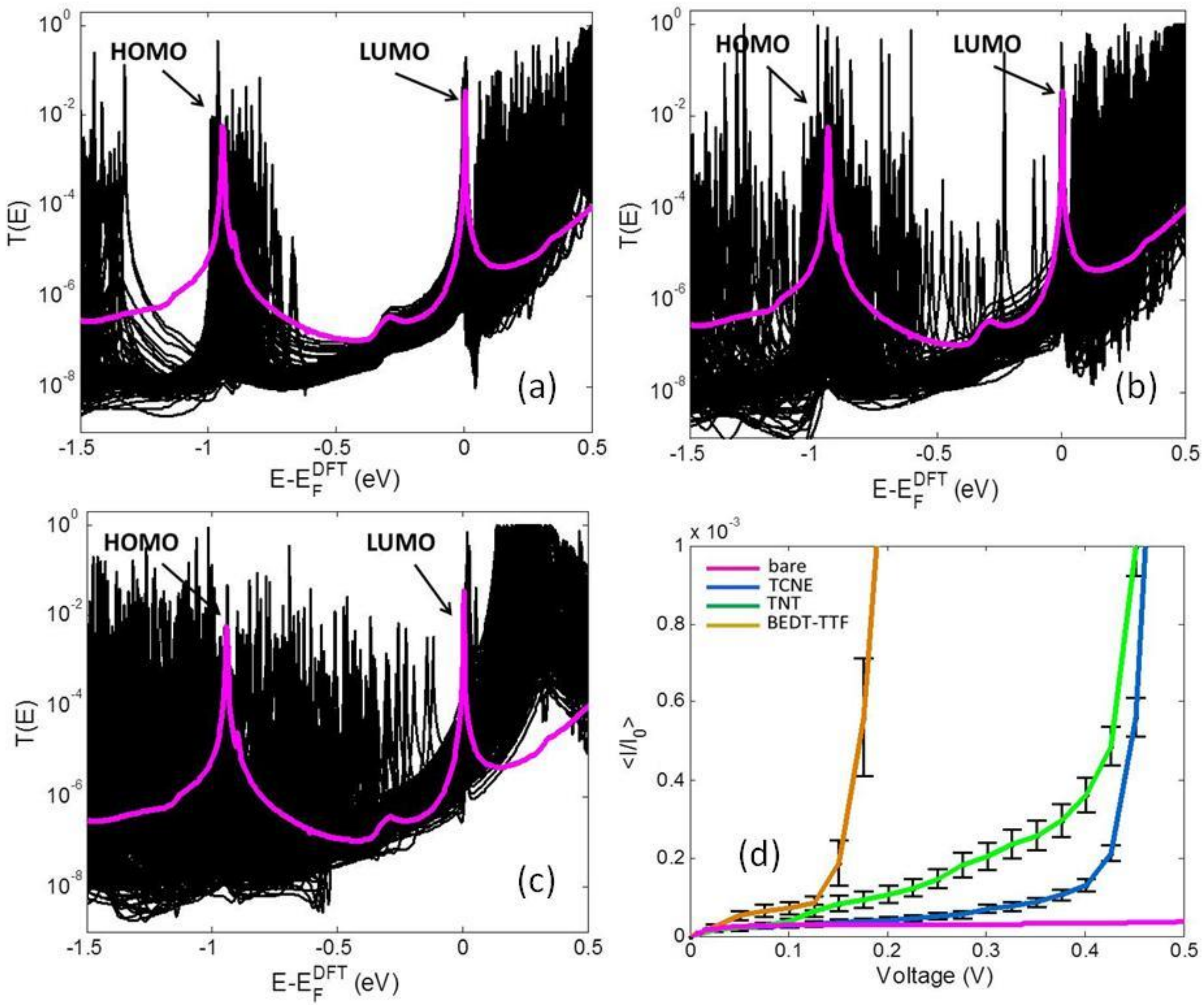

**Figure S5.** DFT calculations of the transmission coefficients as a function of energy for Py-PBI with different configurations of the single-molecules absorbates (a) TCNE , (b) TNT, and (c) BEDT-TTF, and (the pink line) is bare case.. The fourth figure (d) shows the average of current as a function of voltage at T=300K which pass across Py-PBI with three analytes molecules (TCNE, TNT, and BEDT-TTF), where the error bars in figure S5d shows the standard deviation in the means of the currents.

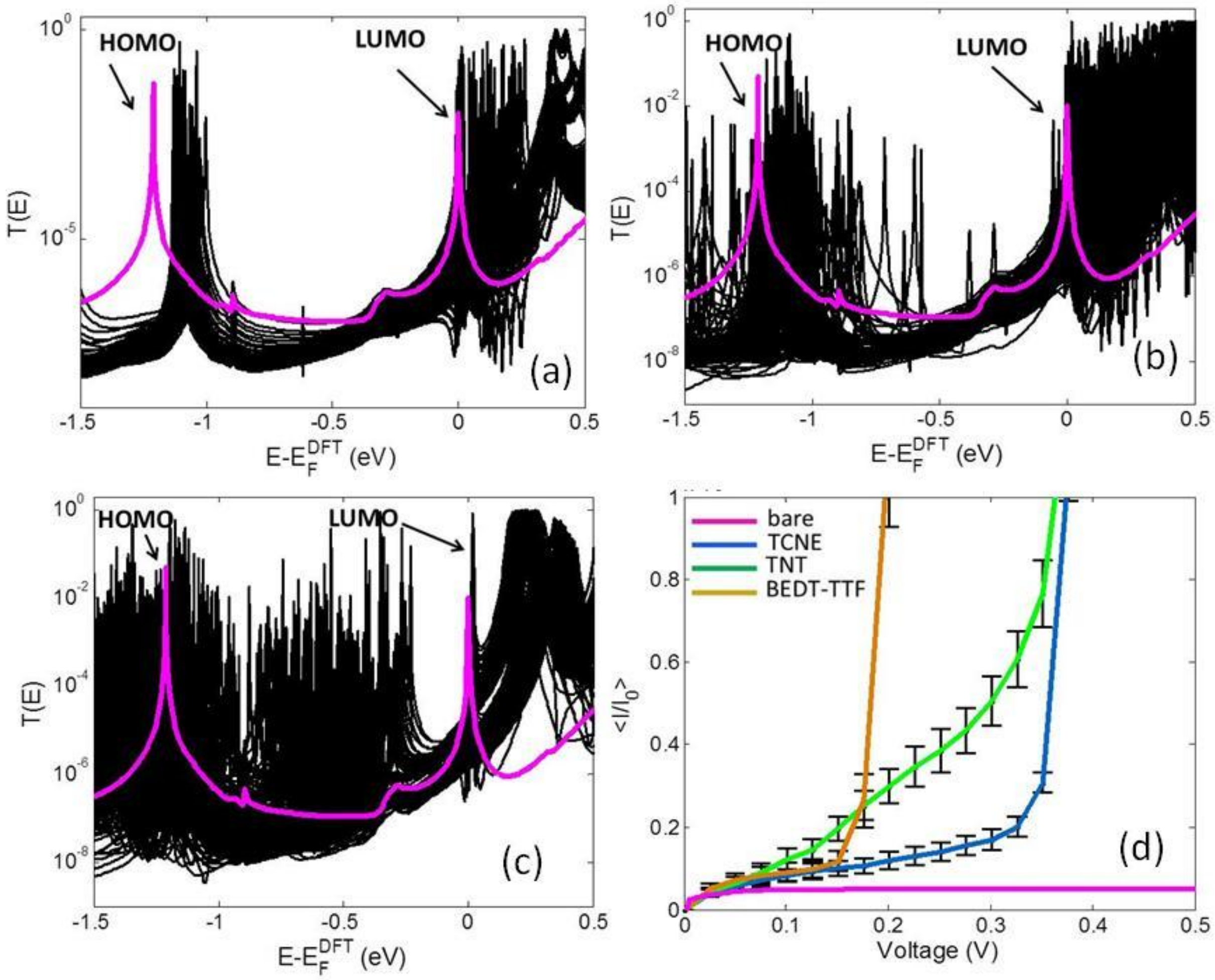

**Figure S6.** DFT calculations of the transmission coefficients as a function of energy for P-PBI with different configurations of the single-molecules absorbates (a) TCNE , (b) TNT, and (c) BEDT-TTF, and (the pink line) is bare case.. The fourth figure (d) shows the average of current as a function of voltage at T=300K which pass across P-PBI with three analytes molecules (TCNE, TNT, and BEDT-TTF), where the error bars in figure S6d shows the standard deviation in the means of the currents.

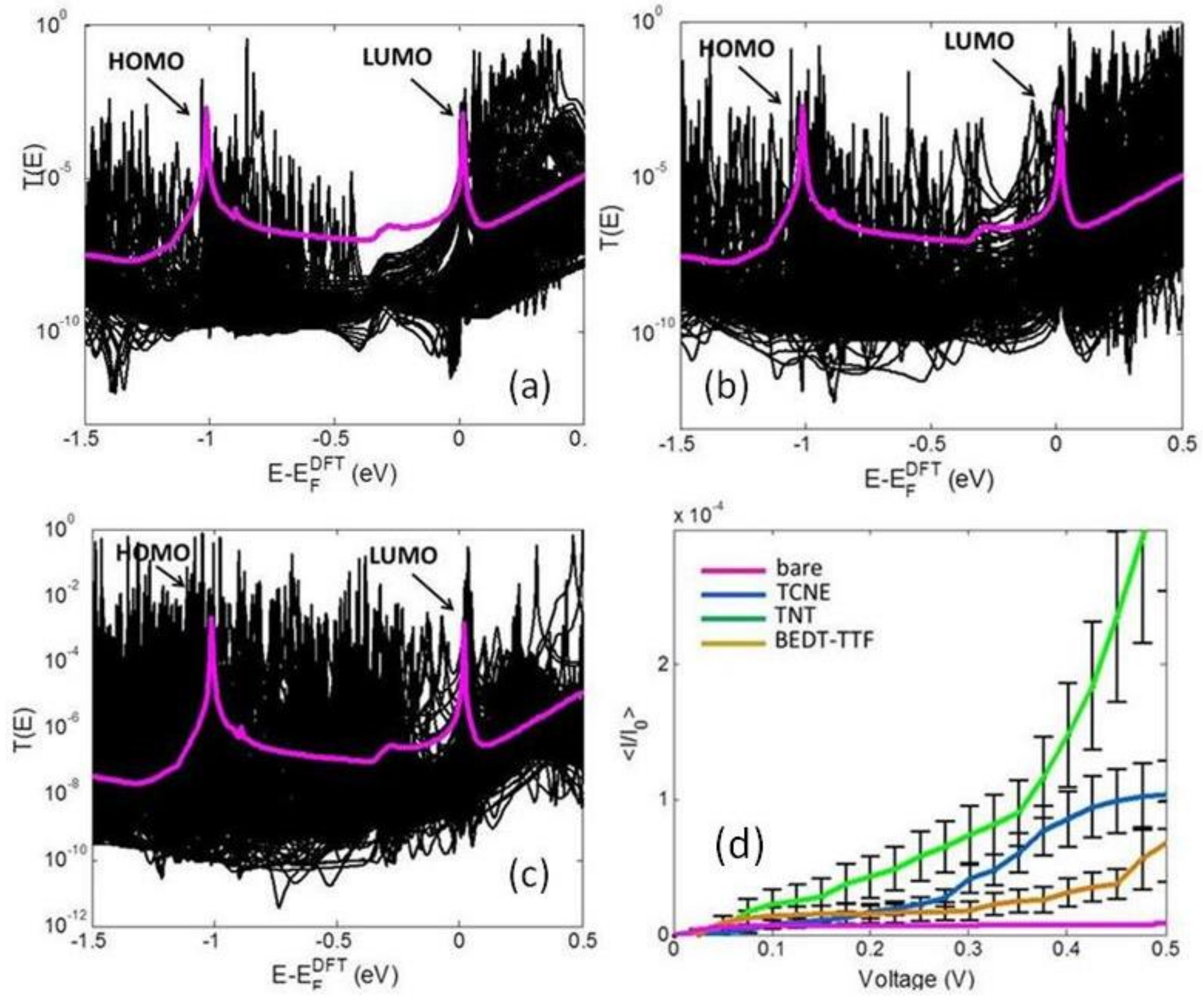

**Figure S7.** DFT calculations of the transmission coefficients as a function of energy for aPy-PBI with different configurations of the single-molecules absorbates (a) TCNE , (b) TNT, and (c) BEDT-TTF, and (the pink line) is bare case.. The fourth figure (d) shows the average of current as a function of voltage at T=300K which pass across aPy-PBI with three analytes molecules (TCNE, TNT, and BEDT-TTF), where the error bars in figure S7d shows the standard deviation in the means of the currents.

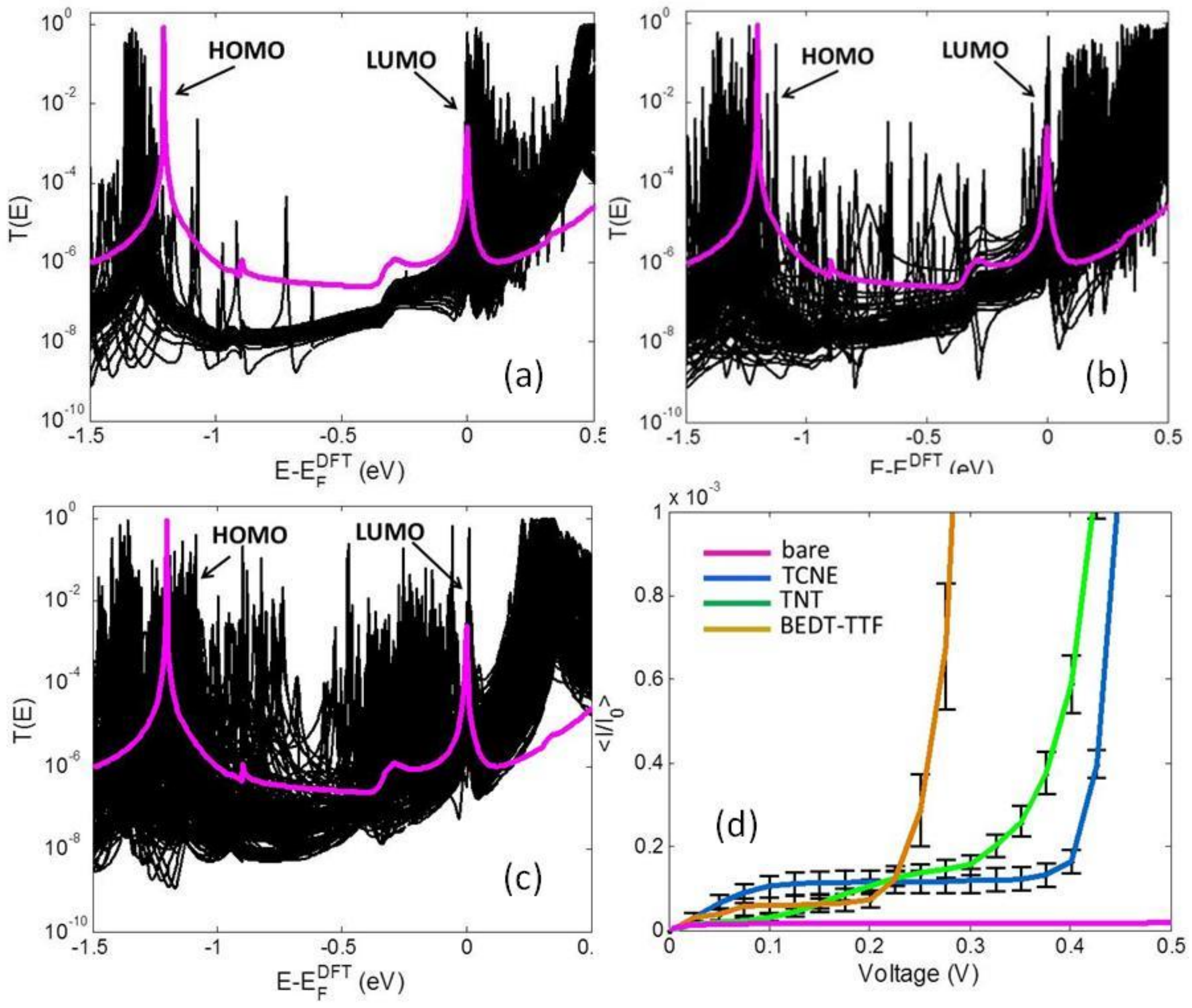

**Figure S8.** DFT calculations of the transmission coefficients as a function of energy for Cl-PBI with different configurations of the single-molecules absorbates (a) TCNE , (b) TNT, and (c) BEDT-TTF, and (the pink line) is bare case.. The fourth figure (d) shows the average of current as a function of voltage at T=300K which pass across Cl-PBI with three analytes molecules (TCNE, TNT, and BEDT-TTF), where the error bars in figure S8d shows the standard deviation in the means of the currents.

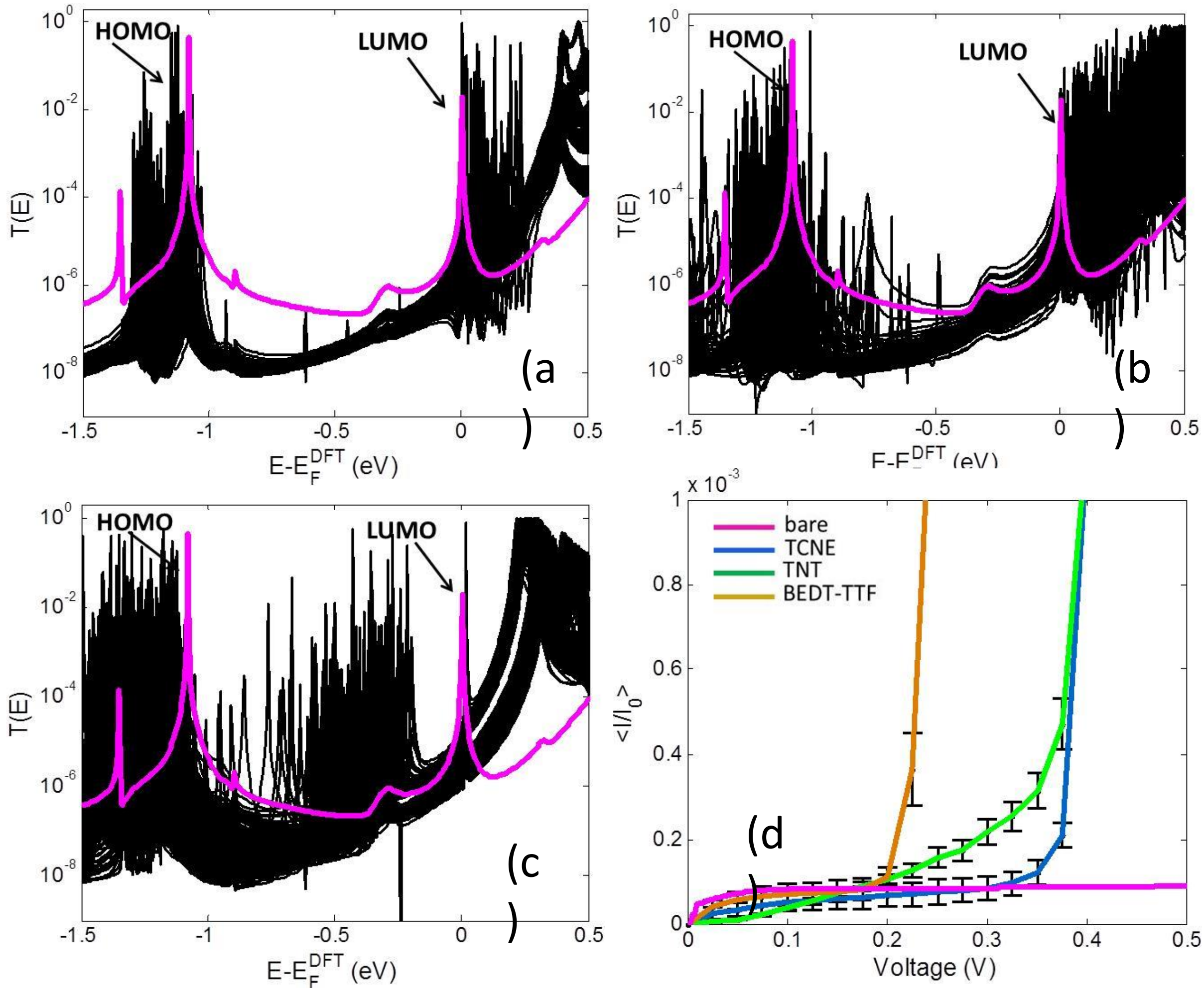

**Figure S9.** DFT calculations of the transmission coefficients as a function of energy for S-PBI with different configurations of the single-molecules absorbates (a) TCNE , (b) TNT, and (c) BEDT-TTF, and (the pink line) is bare case. The fourth figure (d) shows the average of current as a function of voltage at T=300K which pass across S-PBI with three analytes molecules (TCNE, TNT, and BEDT-TTF), where the error bars in figure S9d shows the standard deviation in the means of the currents.

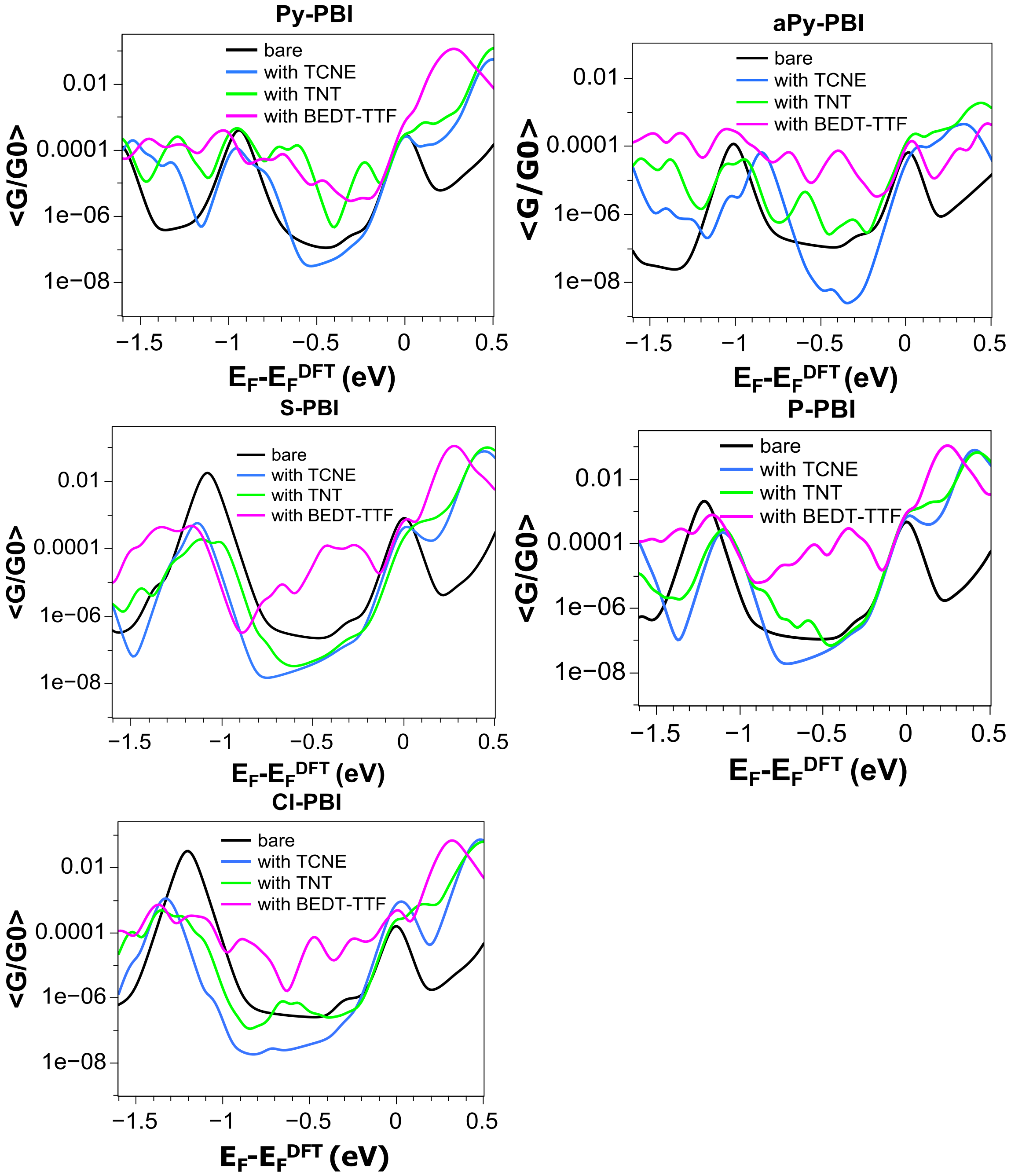

**Figure S10.** The room-temperature, configurationally-averaged as a function of Fermi energy for three analyte molecules (TCNE, TNT, and BEDT-TTF) absorbed on five PBI molecules.

## Locating the optimal value of $E_F$

The optimal value of $E_F$ was chosen by minimising the quantity

$$Y^2(E_F) = \sum_{i=1}^{5} \left( Log(G_i^{theo}) - Log(G_i^{exp}) \right)^2 \qquad (S2)$$

Where $G_i^{theo}$ is the theoretical conductance for a given $E_F$ and $G_i^{exp}$ is the experimental conductance reported in ref [21], where $i$ labels the PBIs. This mean-square deviation between theory and experiment is plotted in figure S8 and shows a minimum at $E_F$ = 0.08eV, which is the values chosen throughout this paper.

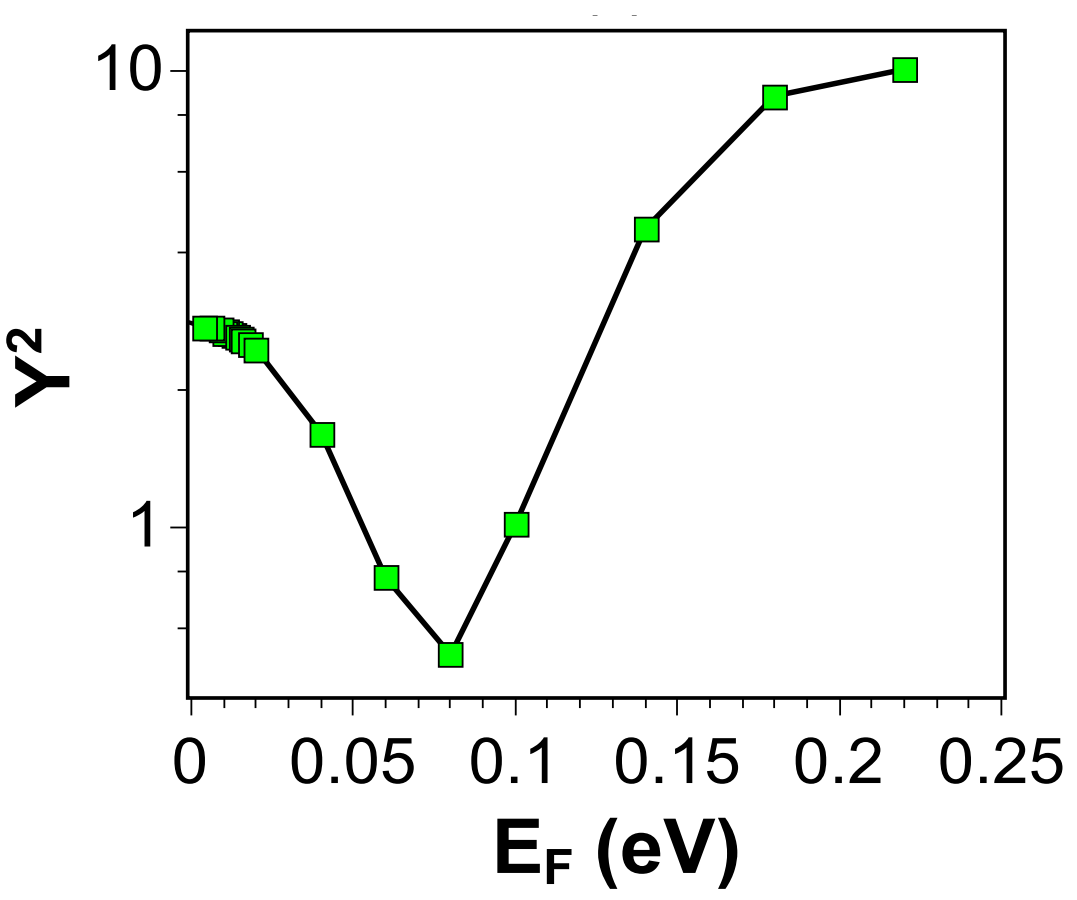

**Figure S11.** The mean square deviation of theory from the experiment as a function of Fermi energy.

**Evaluating of  error in the currents for a distribution of geometries**

The error value of the 214 current curves is given by

$$\sigma_m = \sigma/\sqrt{214 - 1} \qquad (S3)$$

$$\sigma = \sqrt{\frac{1}{214}\sum_{i=1}^{214}(I_i - \langle I \rangle)^2} \qquad (S4)$$

where $\sigma_m$ is standard deviation in the mean of current $\langle I \rangle$.

## Classification of the amount of charge-transfer complex

In particular we used the customised basis set definitions to investigate the effects on the Mulliken population count in SIESTA when using the generalized gradient approximation (GGA-PBE) for the exchange and correlation (GGA) [1], but we also made these calculations with the local density approximation (LDA-CA) [2] and van der waals interactions. The Hamiltonian and overlap matrices are calculated on a real-space grid defined by a plane-wave cutoff of 150 Ry. Each PBI molecule with three analytes which are (TCNE, BEDT-TTF, and TNT) are relaxed into the optimum geometry until the forces on the atoms are smaller than 0.02 eV/Å. Tolerance of Density Matrix is $10^{-4}$ , and in case of the isolated molecules a sufficiently-large unit cell was used. For steric and electrostatic reasons.

**Table S1.** shows DFT calculation of charge- transfer complex and binding energy of TCNE which is analyzed around the backbone of five PBI molecules where all configurations were found in optimum position among 214 configurations for each PBI molecule.

| | Py+TCNE | | aPy+TCNE | | S+TCNE | | P+TCNE | | Cl+TCNE | |
|---|---|---|---|---|---|---|---|---|---|---|
| GGA | 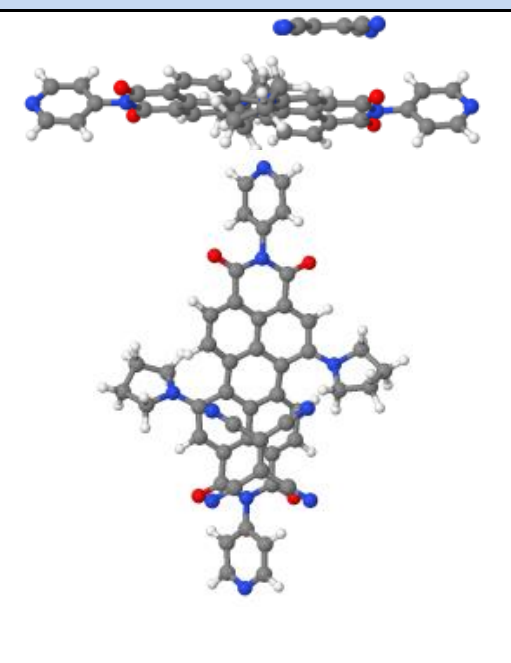 | | 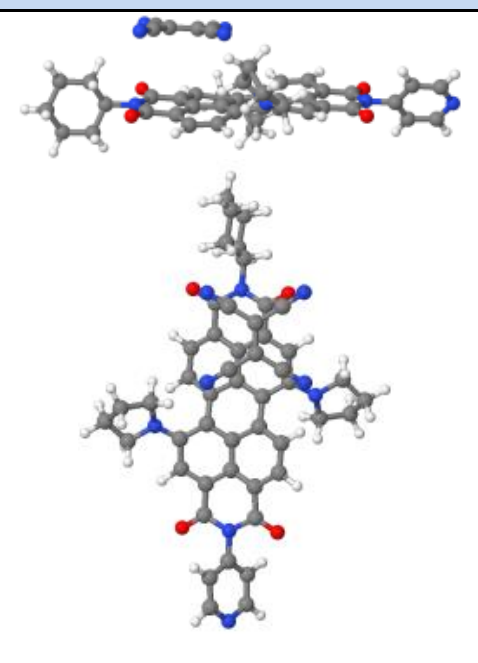 | | 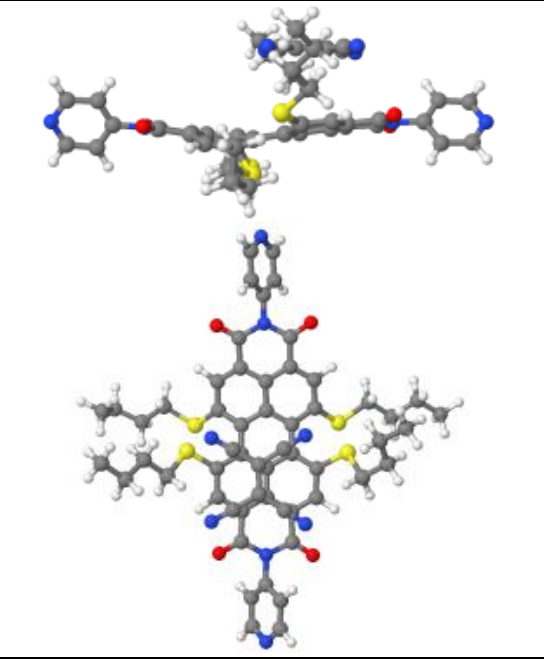 | | 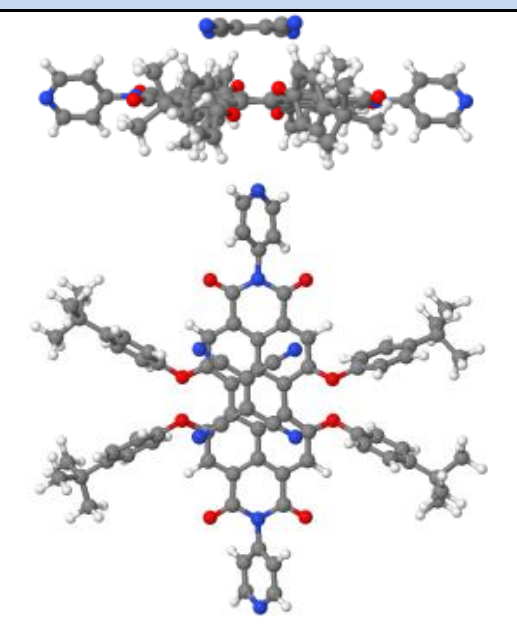 | | 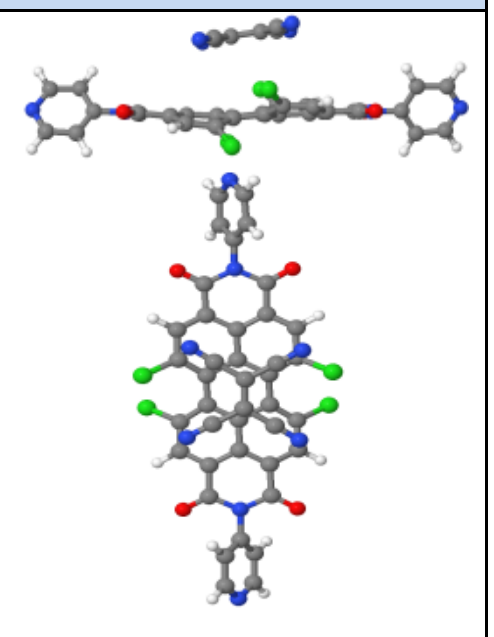 | |
| | ΔQ | ΔE(eV) | ΔQ | ΔE(eV) | ΔQ | ΔE(eV) | ΔQ | ΔE(eV) | ΔQ | ΔE(eV) |
| | 0.311 | -0.301 | 0.317 | -0.293 | 0.279 | -0.294 | 0.292 | -0.245 | 0.289 | -0.027 |
| LDA | 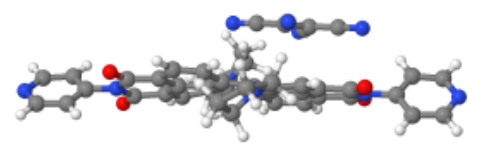 | | 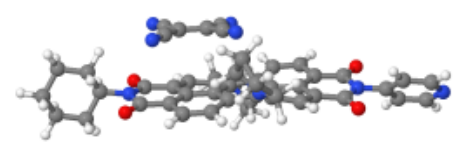 | | 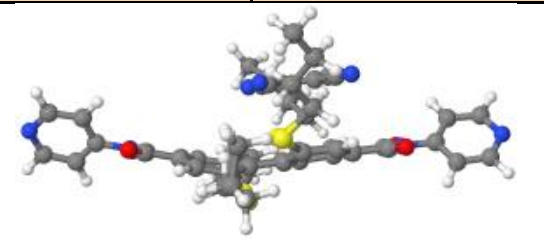 | | 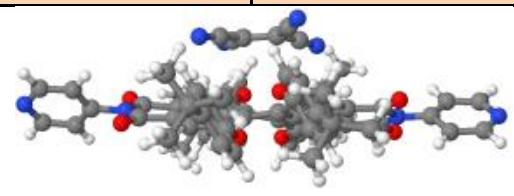 | | 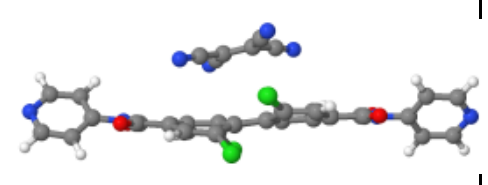 | |
| | ΔQ | ΔE(eV) | ΔQ | ΔE(eV) | ΔQ | ΔE(eV) | ΔQ | ΔE(eV) | ΔQ | ΔE(eV) |
| | 0.32 | -0.955 | 0.331 | -1.021 | 0.295 | -1.075 | 0.298 | -0.783 | 0.293 | -0.477 |
| Vdw | 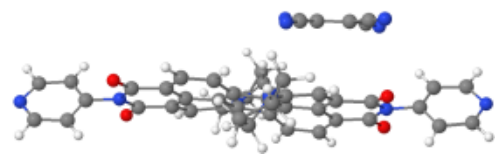 | | 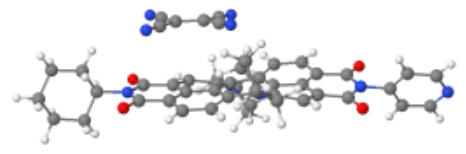 | | 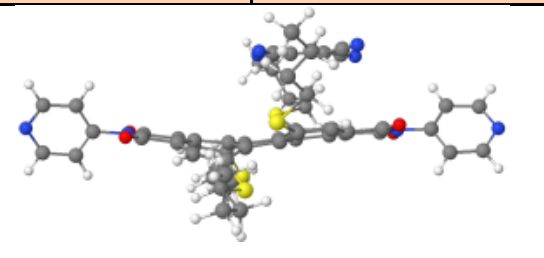 | | 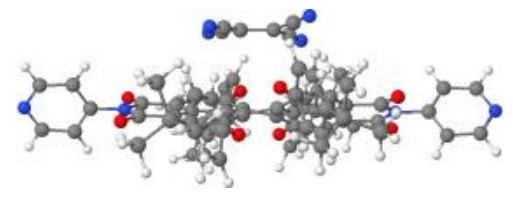 | | 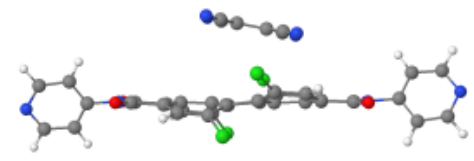 | |
| | ΔQ | ΔE(eV) | ΔQ | ΔE(eV) | ΔQ | ΔE(eV) | ΔQ | ΔE(eV) | ΔQ | ΔE(eV) |
| | 0.307 | -0.7166 | 0.342 | -1.0281 | 0.279 | -0.8636 | 0.294 | -0.8643 | 0.08 | -0.4054 |

**Table S2.** shows DFT calculation of charge- transfer complex and binding energy of BEDT-TTF which is analyzed around the backbone of five PBI molecules where all configurations were found in optimum position among 214 configurations for each PBI molecule.

| | Py+BEDT-TTF | | aPy+ BEDT-TTF | | S+ BEDT-TTF | | P+ BEDT-TTF | | Cl+ BEDT-TTF | |
|---|---|---|---|---|---|---|---|---|---|---|
| GGA | 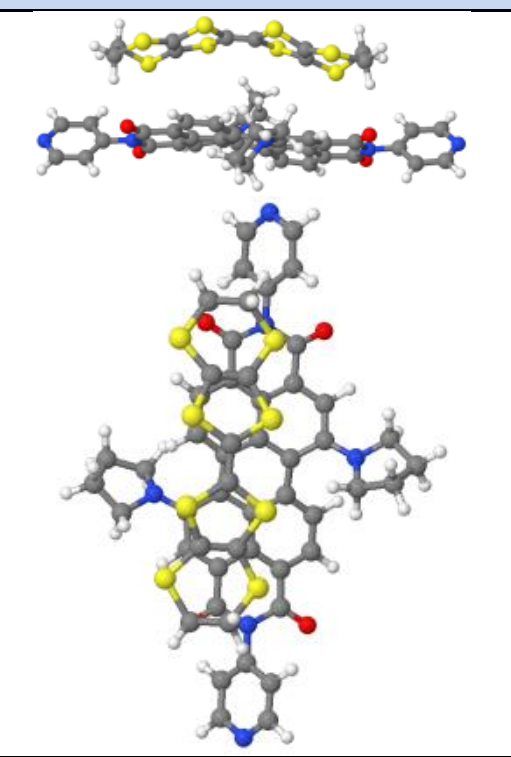 | | 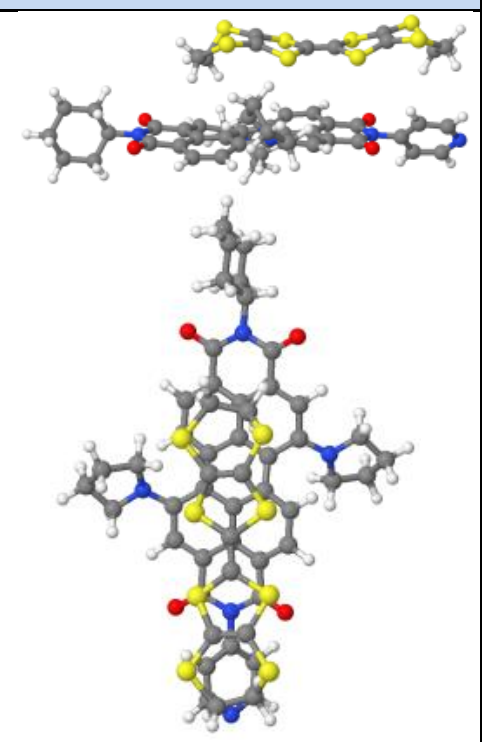 | | 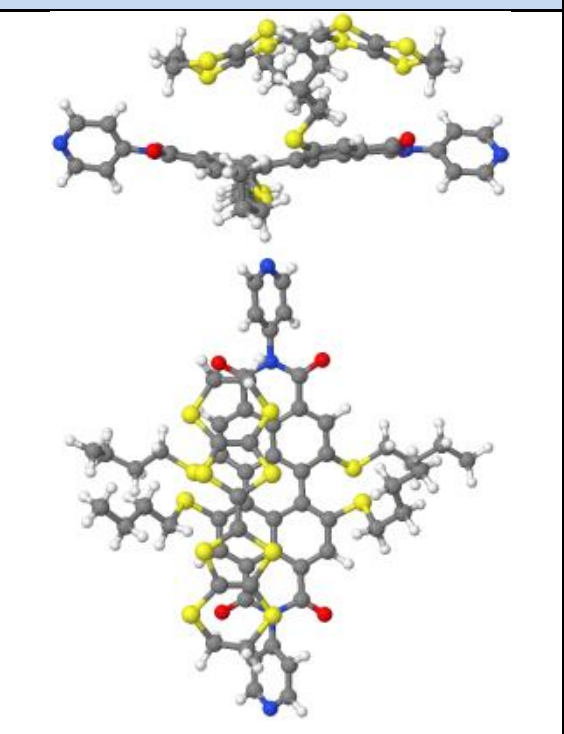 | | 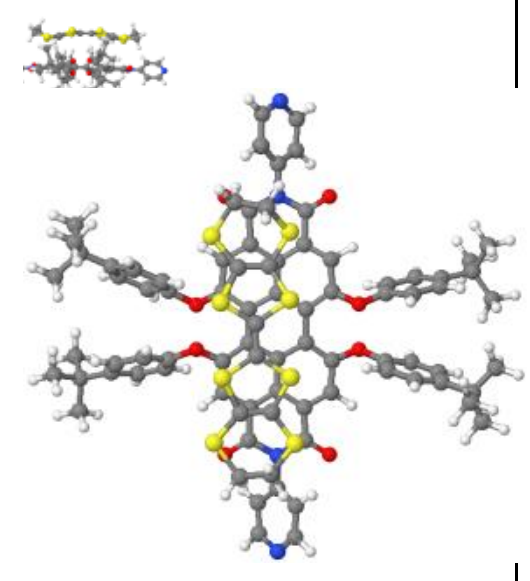 | | 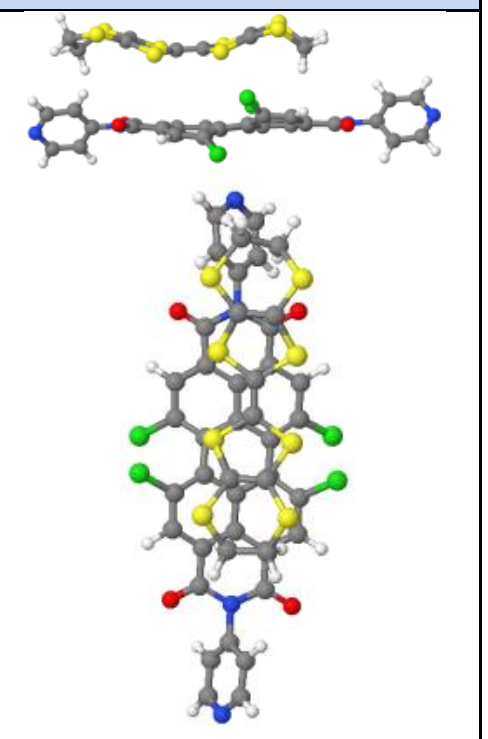 | |
| | ΔQ | ΔE(eV) | ΔQ | ΔE(eV) | ΔQ | ΔE(eV) | ΔQ | ΔE(eV) | ΔQ | ΔE(eV) |
| | -0.043 | -0.284 | -0.053 | -0.199 | -0.007 | -0.225 | -0.179 | -0.151 | -0.239 | -0.1820 |
| LDA | 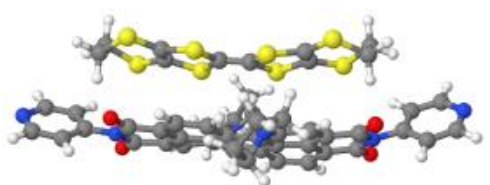 | | 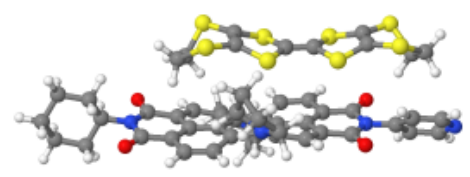 | | 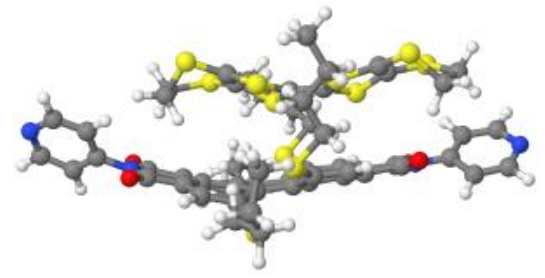 | | 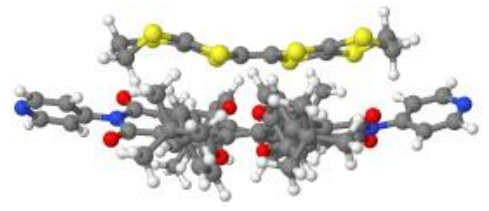 | | 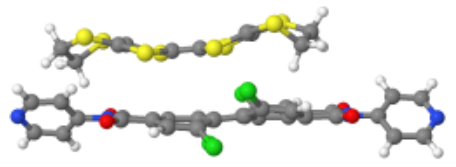 | |
| | ΔQ | ΔE(eV) | ΔQ | ΔE(eV) | ΔQ | ΔE(eV) | ΔQ | ΔE(eV) | ΔQ | ΔE(eV) |
| | -0.222 | -1.236 | -0.148 | -0.991 | -0.104 | -1.2 | -0.214 | -1.2 | -0.299 | -1.0837 |
| VdW | 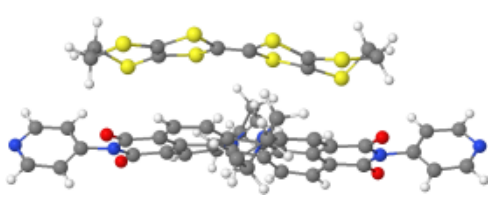 | | 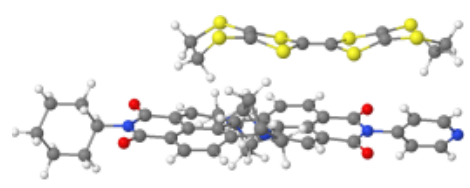 | | 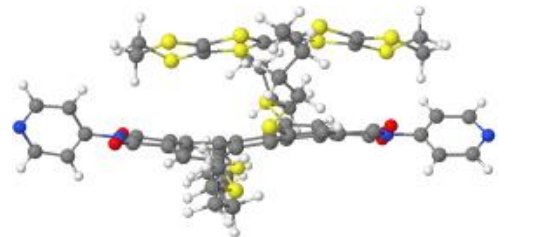 | | 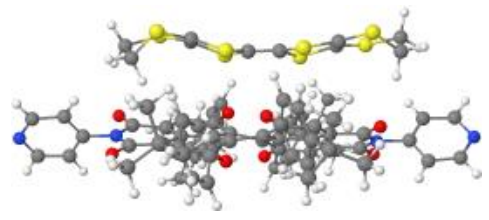 | | 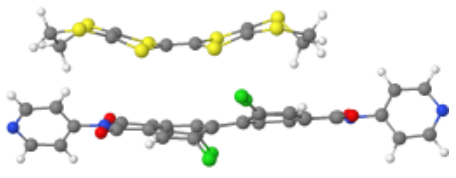 | |
| | ΔQ | ΔE(eV) | ΔQ | ΔE(eV) | ΔQ | ΔE(eV) | ΔQ | ΔE(eV) | ΔQ | ΔE(eV) |
| | -0.166 | -1.184 | -0.079 | -1.0064 | -0.064 | -1.1611 | -0.133 | -1.1837 | -0.224 | -1.0432 |

**Table S3.** shows DFT calculation of charge- transfer complex and binding energy of TNT which is analyzed around the backbone of five PBI molecules where all configurations were found in optimum position among 214 configurations for each PBI molecule.

| | Py+TNT | | aPy+TNT | | S+TNT | | P+TNT | | Cl+TNT | |
|---|---|---|---|---|---|---|---|---|---|---|
| GGA | 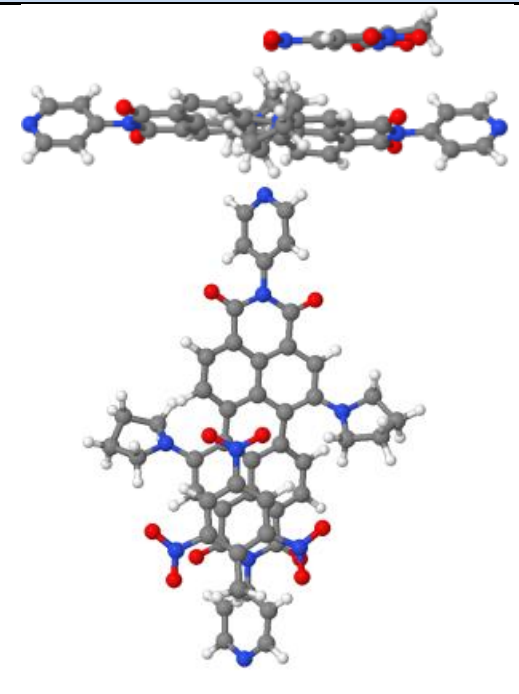 | | 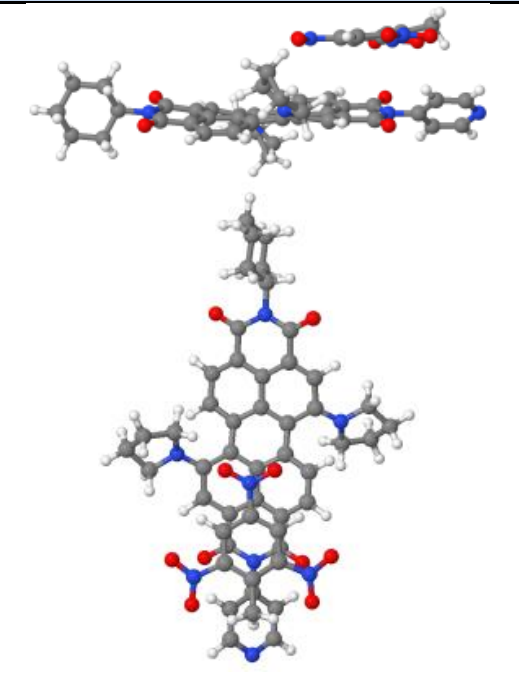 | | 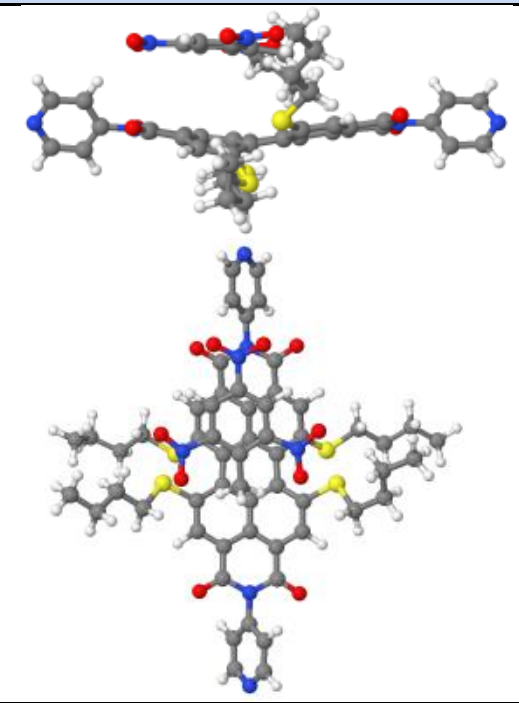 | | 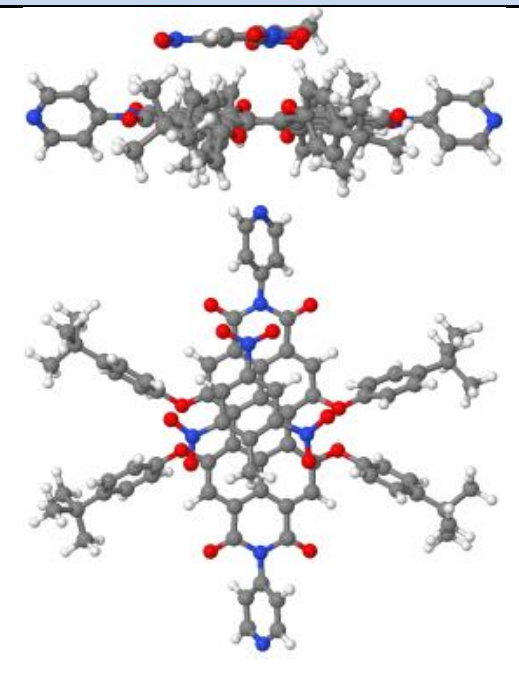 | | 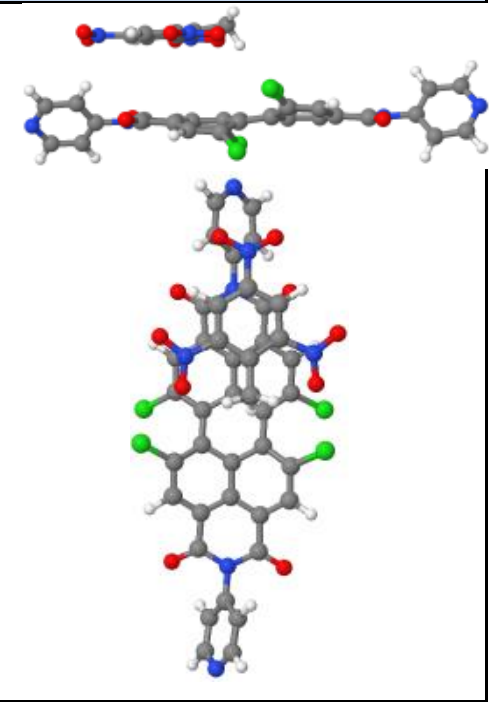 | |
| | ΔQ | ΔE(eV) | ΔQ | ΔE(eV) | ΔQ | ΔE(eV) | ΔQ | ΔE(eV) | ΔQ | ΔE(eV) |
| | 0.023 | -0.141 | 0.11 | -0.097 | 0.132 | -0.102 | 0.049 | -0.0921 | 0.011 | -0.015 |
| LDA | 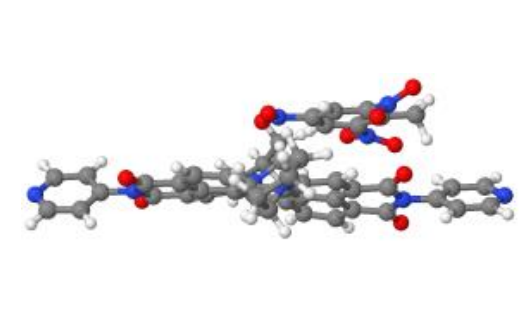 | | 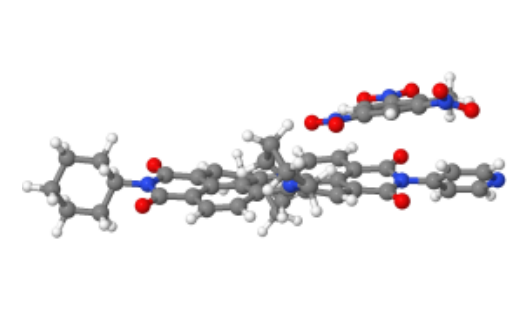 | | 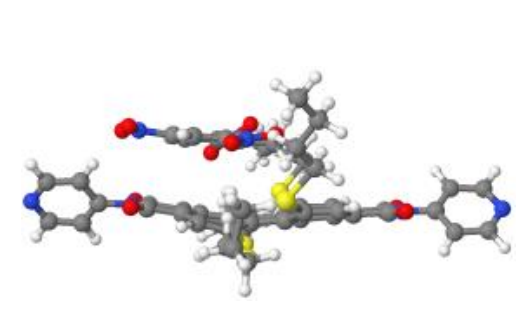 | | 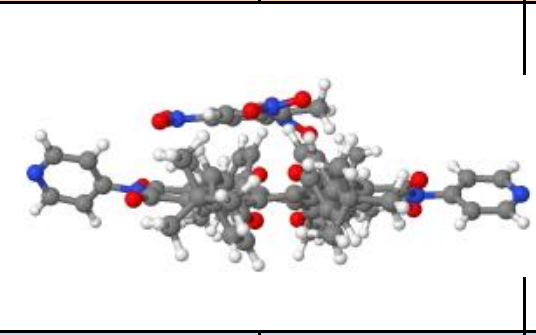 | | 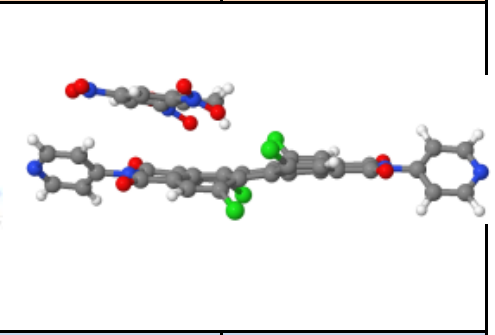 | |
| | ΔQ | ΔE(eV) | ΔQ | ΔE(eV) | ΔQ | ΔE(eV) | ΔQ | ΔE(eV) | ΔQ | ΔE(eV) |
| | 0.028 | -0.813 | 0.12 | -0.833 | 0.331 | -0.8754 | 0.054 | -0.85 | 0.011 | -0.5032 |
| Vdw | 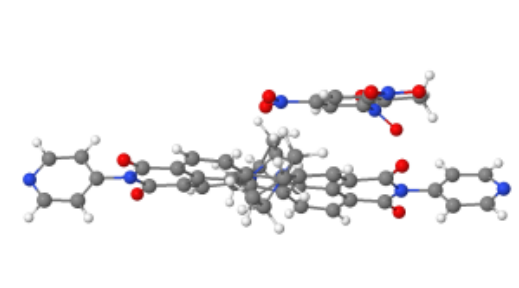 | | 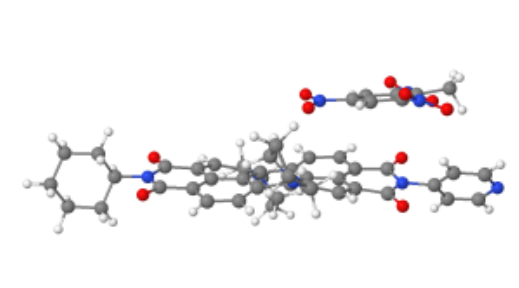 | | 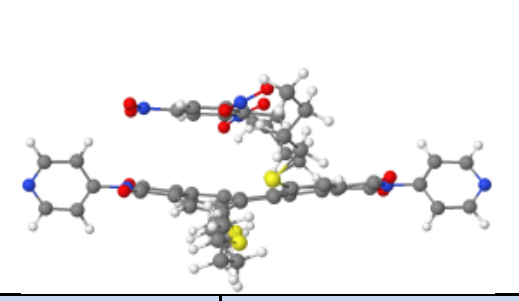 | | 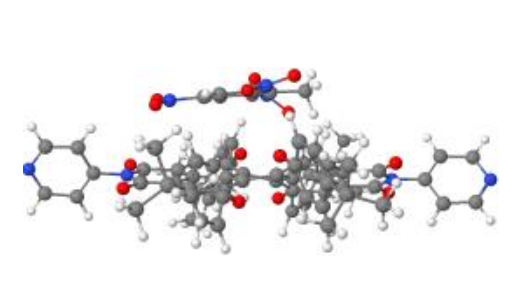 | | 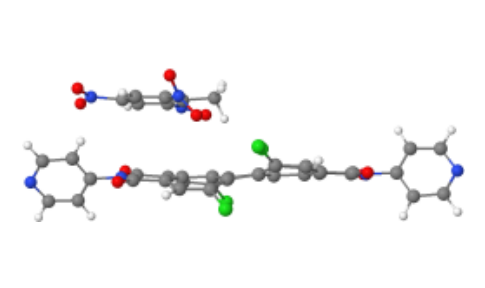 | |
| | ΔQ | ΔE(eV) | ΔQ | ΔE(eV) | ΔQ | ΔE(eV) | ΔQ | ΔE(eV) | ΔQ | ΔE(eV) |
| | 0.023 | -0.933 | 0.114 | -0.723 | 0.048 | -0.966 | 0.051 | -1.082 | 0.024 | -0.614 |

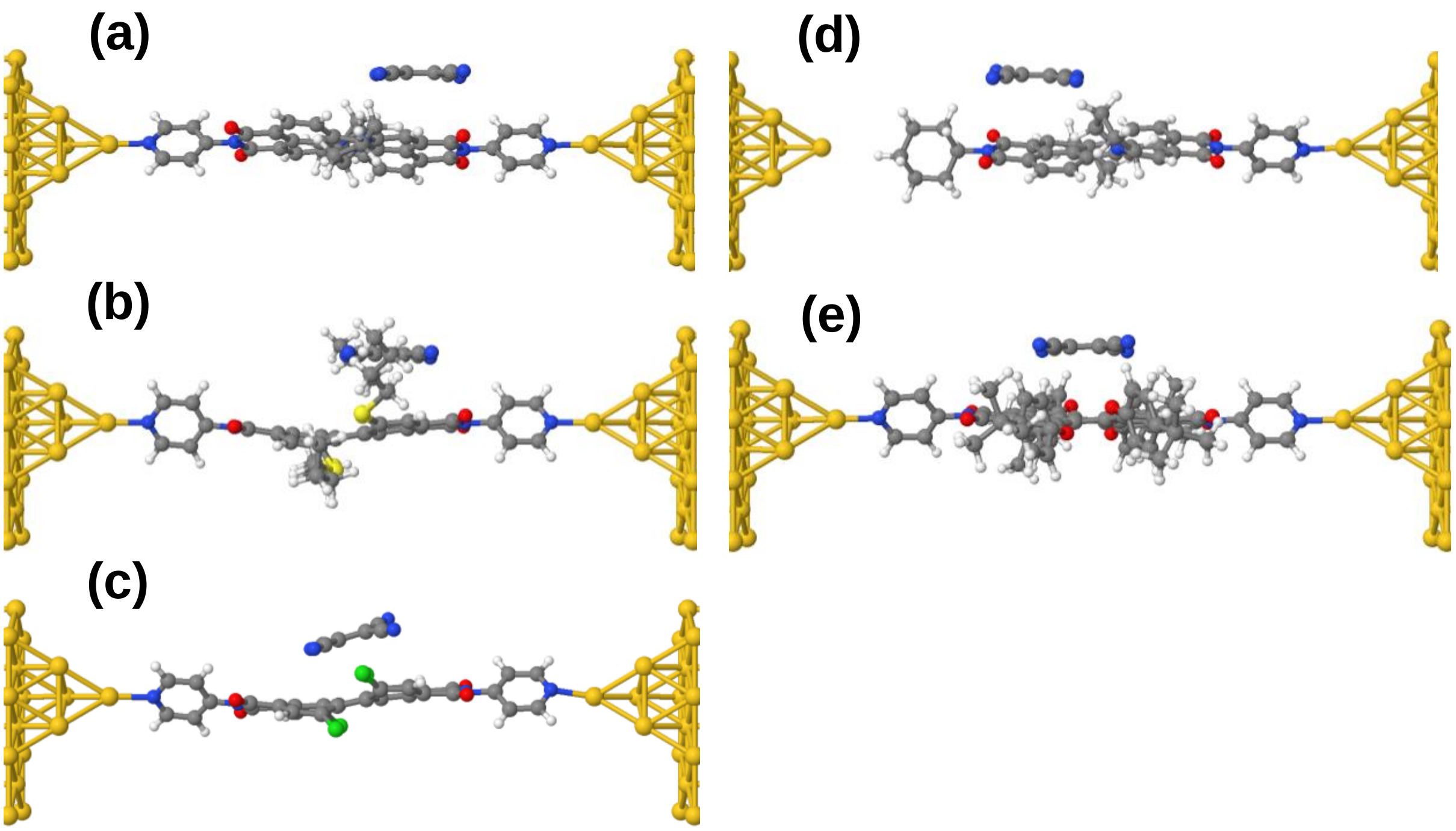

**Figure S12. O**ptimized configurations of a single TCNE adsorbed on (a) Py-PBI, (b) aPy-PBI, (c) S-PBI, (d) P-PBI, and (e) Cl-PBI attached to two metallic leads.

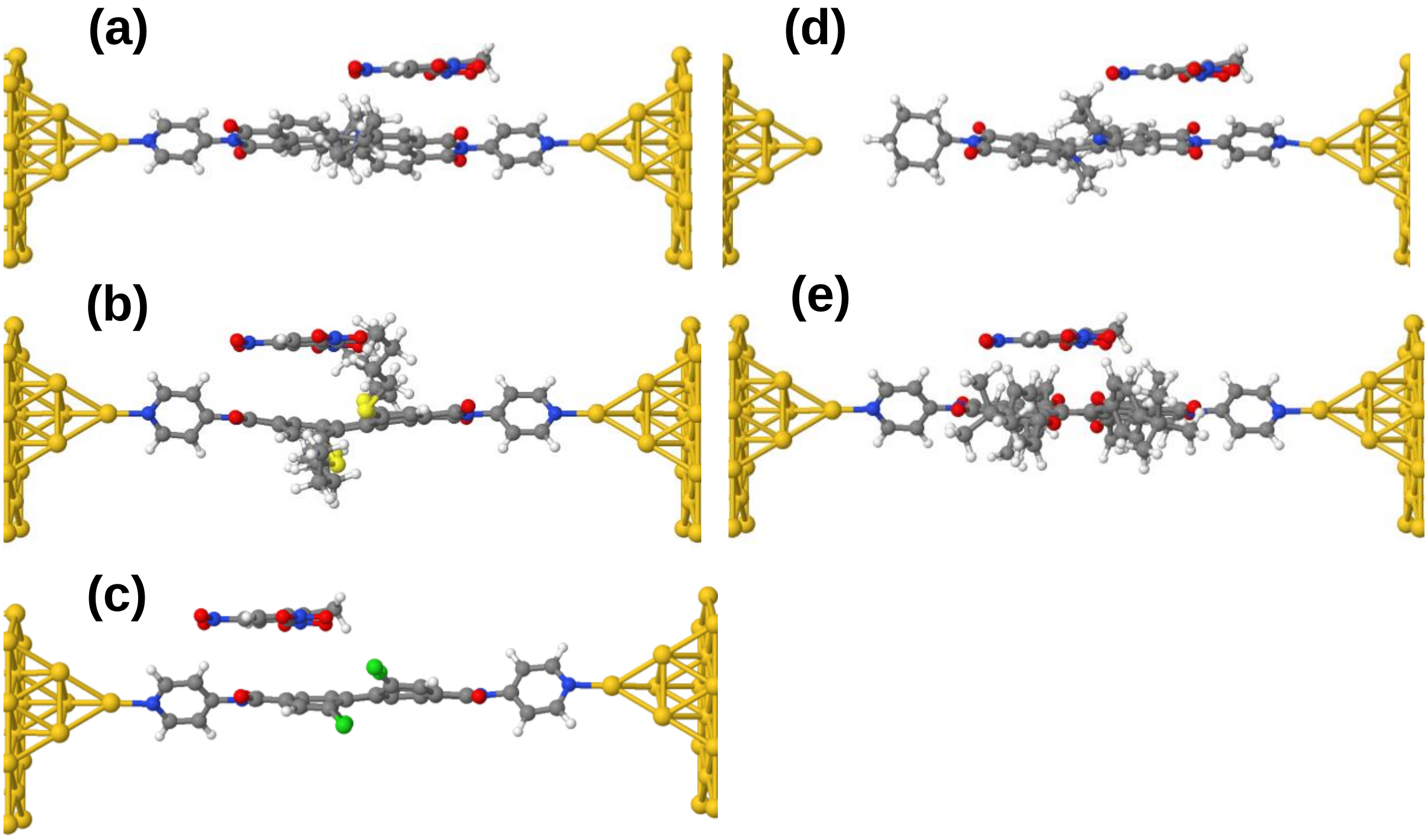

**Figure S13. O**ptimized configurations of a single TNT adsorbed on (a) Py-PBI, (b) aPy-PBI, (c) S-PBI, (d) P-PBI, and (e) Cl-PBI attached to two metallic leads.

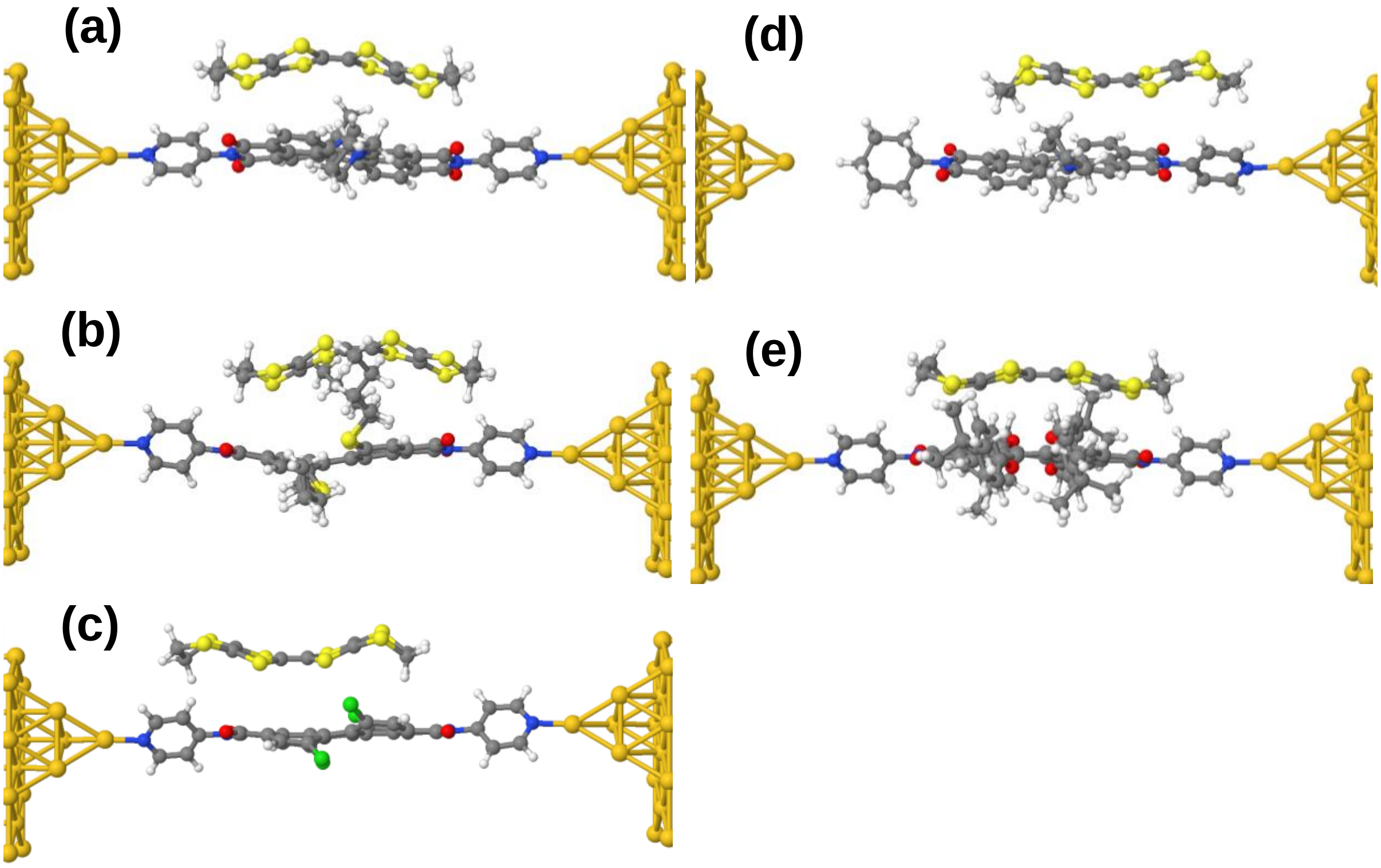

**Figure S14. O**ptimized configurations of a single BEDT-TTF adsorbed on (a) Py-PBI, (b) aPy-PBI, (c) S-PBI, (d) P-PBI, and (e) Cl-PBI attached to two metallic leads.

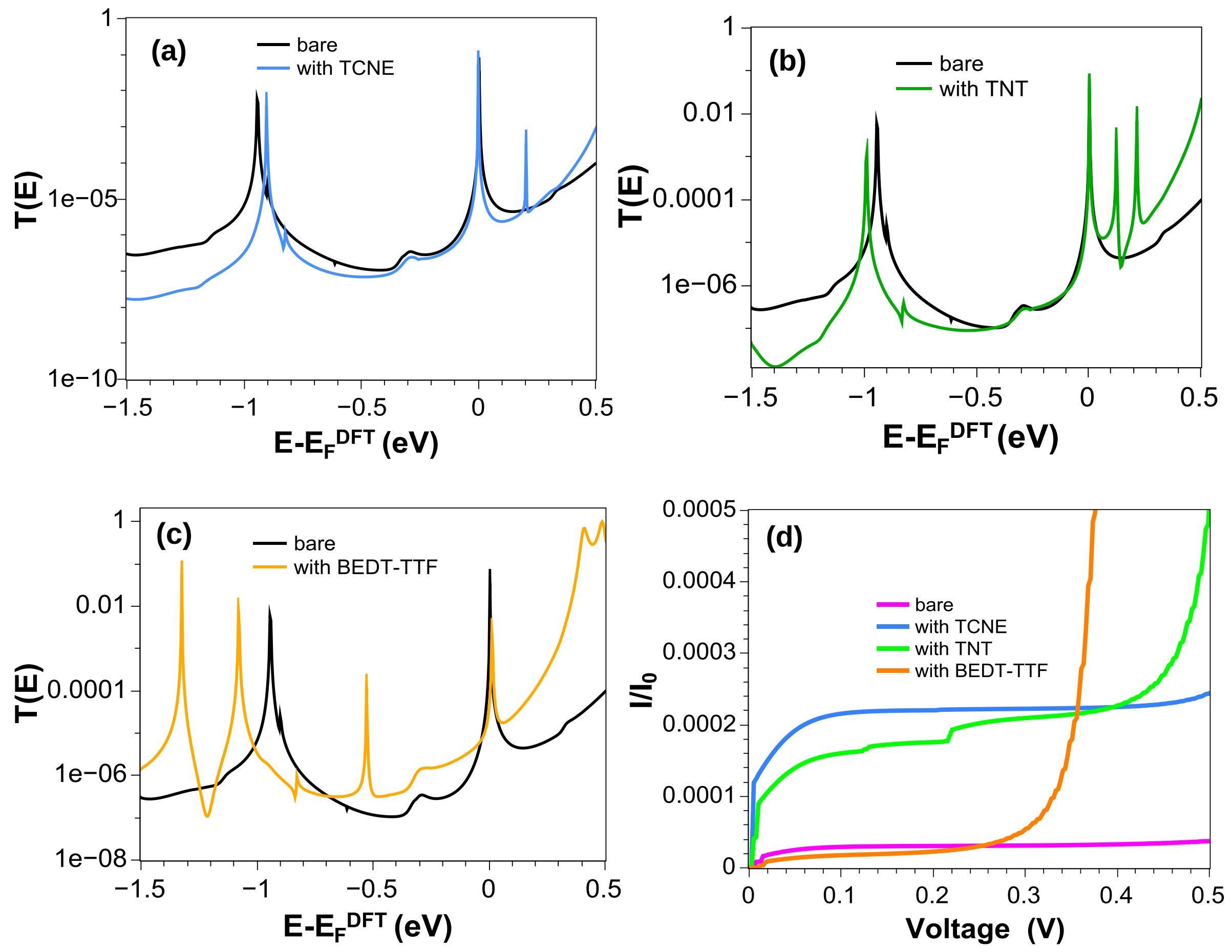

**Figure S15.** DFT calculations using LDA approximation of the transmission coefficients as a function of energy at T=0K for optimum configuration of Py-PBI with (a) TCNE , (b) TNT, and (c) BEDT-TTF. The fourth figure (d) shows the current as a function of voltage at T=300K for the bare Py-PBI and in the presence of the three analyte molecules (TCNE, TNT, and BEDT-TTF).

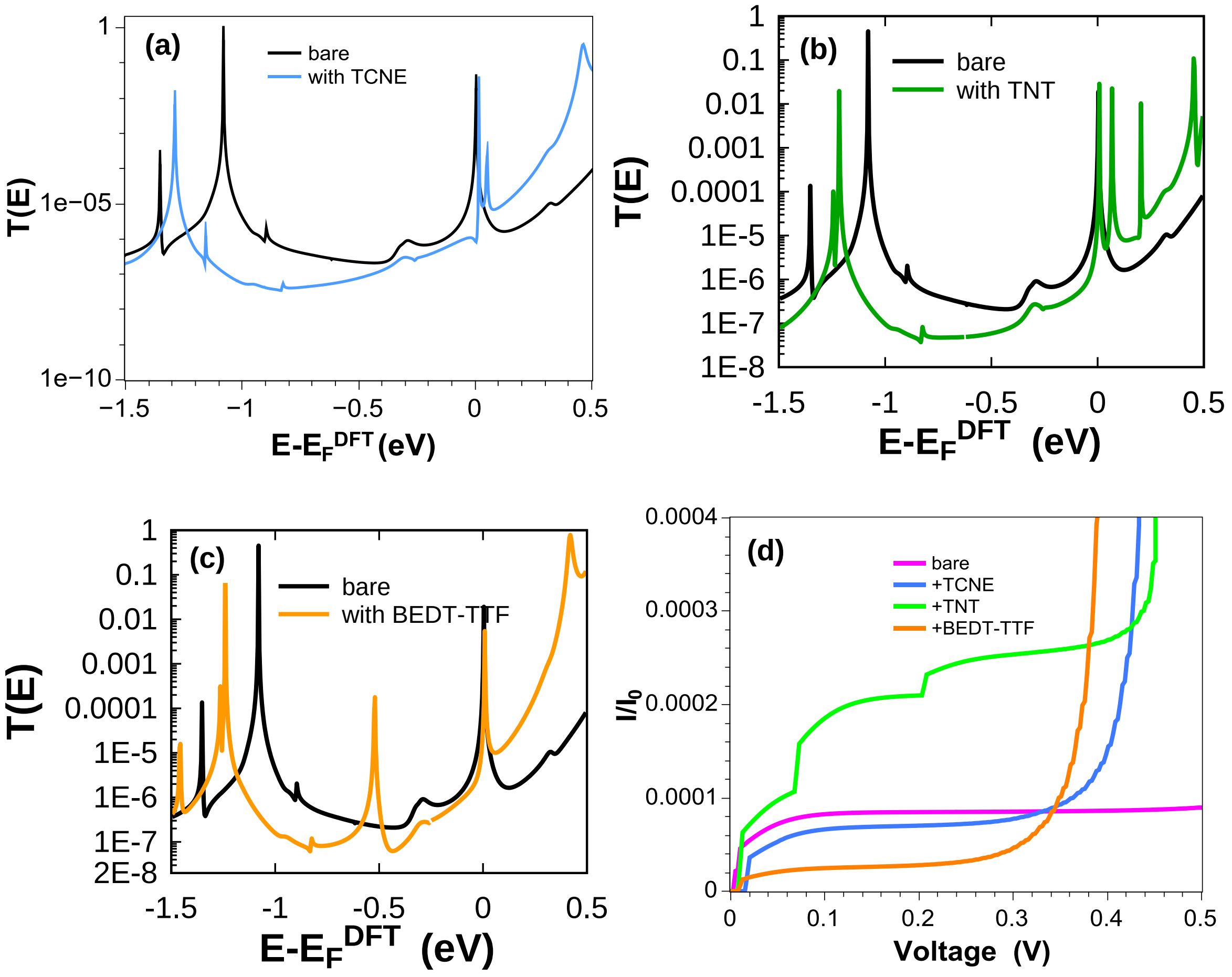

**Figure S16.** DFT calculations using LDA approximation of the transmission coefficients as a function of energy at T=0K for optimum configuration of S-PBI with (a) TCNE , (b) TNT, and (c) BEDT-TTF. The fourth figure (d) shows the current as a function of voltage at T=300K for the bare S-PBI and in the presence of the three analyte molecules (TCNE, TNT, and BEDT-TTF).

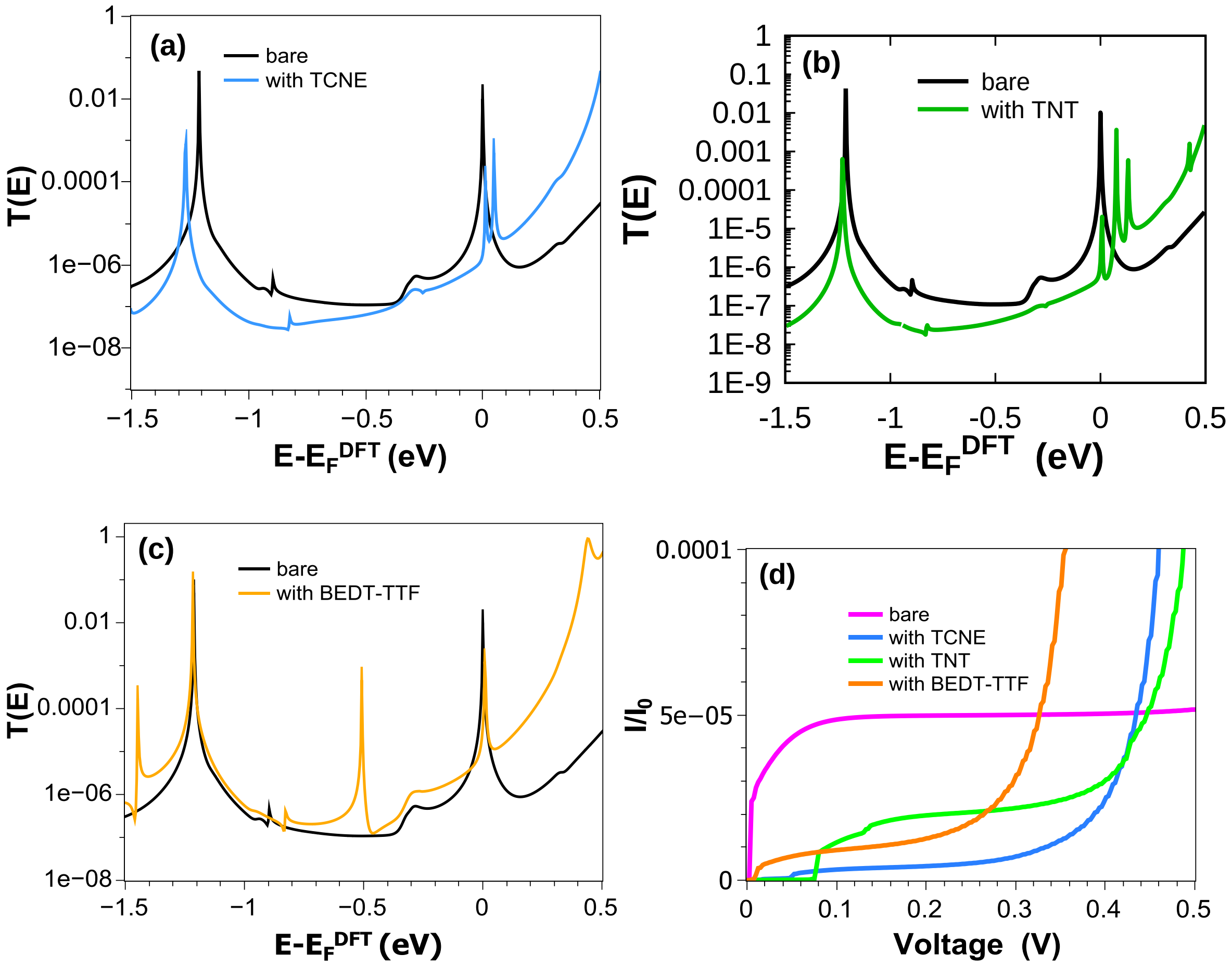

**Figure S17.** DFT calculations using LDA approximation of the transmission coefficients as a function of energy at T=0K for optimum configuration of P-PBI with (a) TCNE , (b) TNT, and (c) BEDT-TTF. The fourth figure (d) shows the current as a function of voltage at T=300K for the bare P-PBI and in the presence of the three analyte molecules (TCNE, TNT, and BEDT-TTF).

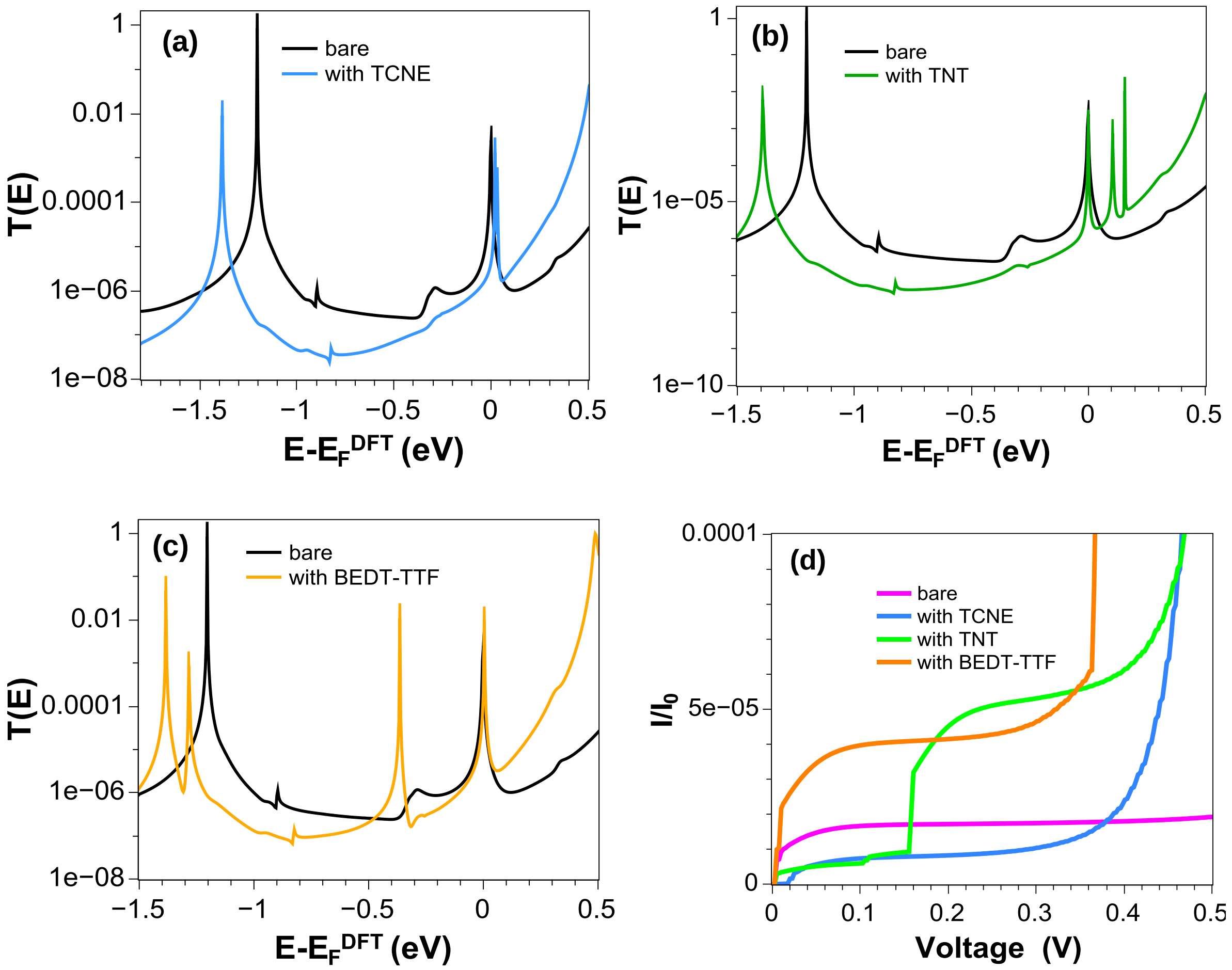

**Figure S18.** DFT calculations using LDA approximation of the transmission coefficients as a function of energy at T=0K for optimum configuration of Cl-PBI with (a) TCNE , (b) TNT, and (c) BEDT-TTF. The fourth figure (d) shows the current as a function of voltage at T=300K for the bare Cl-PBI and in the presence of the three analyte molecules (TCNE, TNT, and BEDT-TTF).

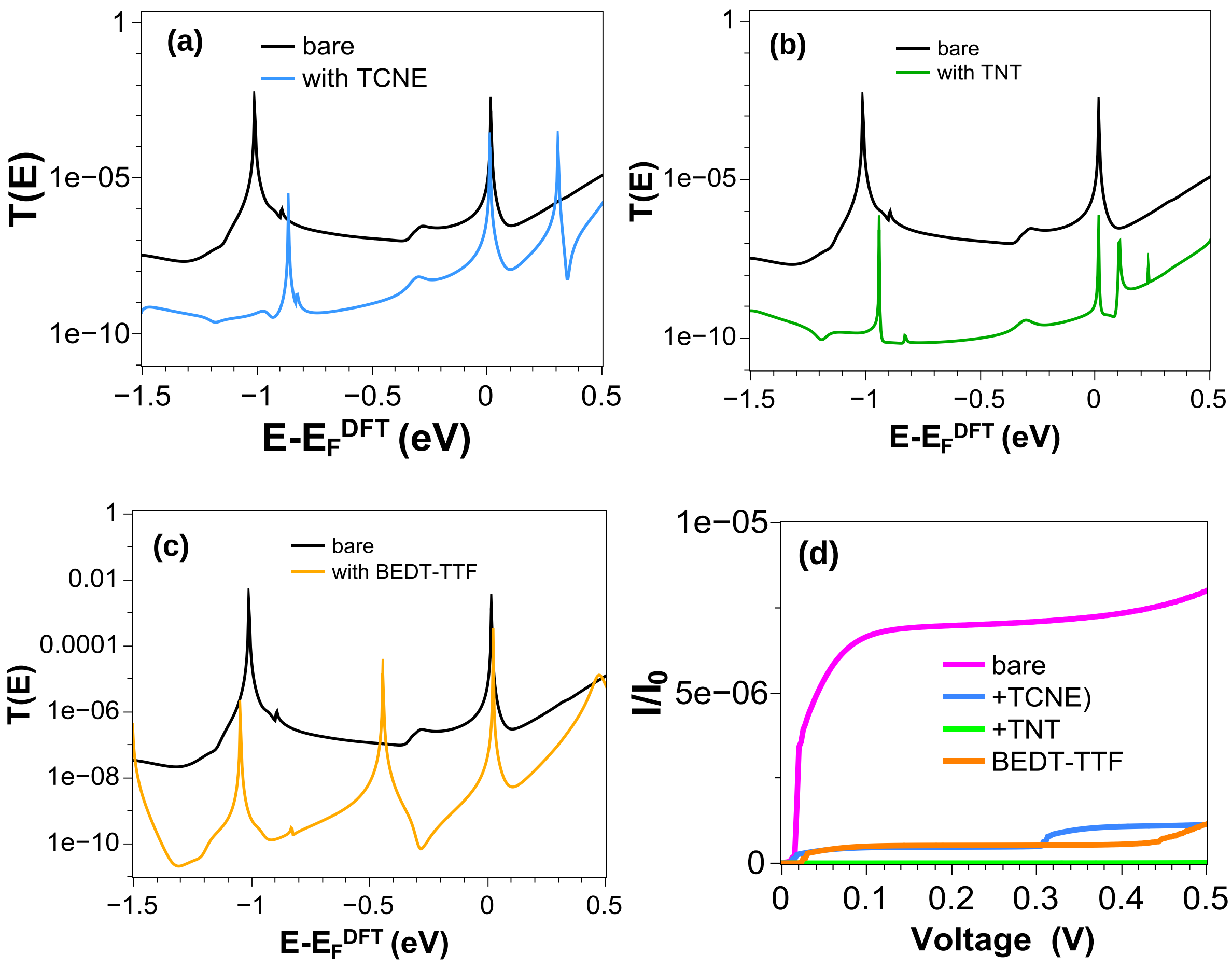

**Figure S19.** DFT calculations using LDA approximation of the transmission coefficients as a function of energy at T=0K for optimum configuration of aPy-PBI with (a) TCNE , (b) TNT, and (c) BEDT-TTF. The fourth figure (d) shows the current as a function of voltage at T=300K for the bare aPy-PBI and in the presence of the three analyte molecules (TCNE, TNT, and BEDT-TTF).

It is interesting to note that the behaviour of the ensemble average is qualitatively differenet from the ensemble average. Figure S19 shows the transmission coefficient for the molecule aPy-PBI, in the absence and presence of the analytes. It is clear that the conductance in the presence of the analytes in their optimal binding configurations is lower than in the case of bare molecule. In contrast, figure S7 shows that the ensemble-averaged conductance, is higher in the presence of the analytes.

**Quantifying the sensitivity of the PBIs for discriminating sensing.**

To quantify the potential of the five PBI derivatives (labeled j=1,…, 5) for the discriminating sensing of the three analytes TCNE, TNT and BEDT-TTF, (labeled n=1,2,3) we calculated the ensemble-averaged, room-temperature currents $I_{jn}$ as a function of voltage $V$ and computed the following correlators of the currents

$$A_j^{nm} = \int_{-V/2}^{V/2} dV \left( I_{jn}(V) - I_{jm}(V) \right)^2 \quad \text{(S5)}$$

These were then normalized by the squared currents of the bare backbones to yield the quantities:

$$X_j^{nm} = \frac{A_j^{nm}}{\int_{-V/2}^{V/2} dV \left( I_j(V_{bare}) \right)^2} \quad \text{(S6)}$$

For $n \neq m$, table 1 shows the values of $X_j^{nm}$ obtained for V=1volt. Clearly, as defined, $X_j^{nm} = 0$ when $n = m$. In practice, a sensing event would involve measurement of a new set of curves ($I_{jm}(V)$ in equation 2) and combining these with a 'calibration set of curves ($I_{jn}(V)$ in equation 2), in which case $X_j^{nm}$ would be small but not zero when $n = m$.

**Table 1.** The values of $X_j^{nm}$

| J | PBIs | $X_j^{12}$ | $X_j^{13}$ | $X_j^{23}$ |
|---|---|---|---|---|
| 1 | aPy-PBI | 0.3043 | 0.2776 | 0.0044 |
| 2 | Cl-PBI | 0.0031 | 0.2457 | 0.2508 |
| 3 | P-PBI | 0.0021 | 0.4101 | 0.4011 |
| 4 | Py-PBI | 0.3641 | 0.4365 | 0.9161 |
| 5 | S-PBI | 0.0309 | 0.1344 | 0.1748 |
| SUM | | 0.7045 | 1.5043 | 1.7472 |

Table 1 and figure 14 shows the value of $X_j^{nm}$ when $n \neq m$. Ideally, to avoid false positives, these numbers should be as large as possible and since they are largest for Py-PBI, we conclude that Py-PBI is the best individual sensor. Nevertheless, for the most accurate discriminating sensing, the fingerprint of an analyte across all five backbones should be used.

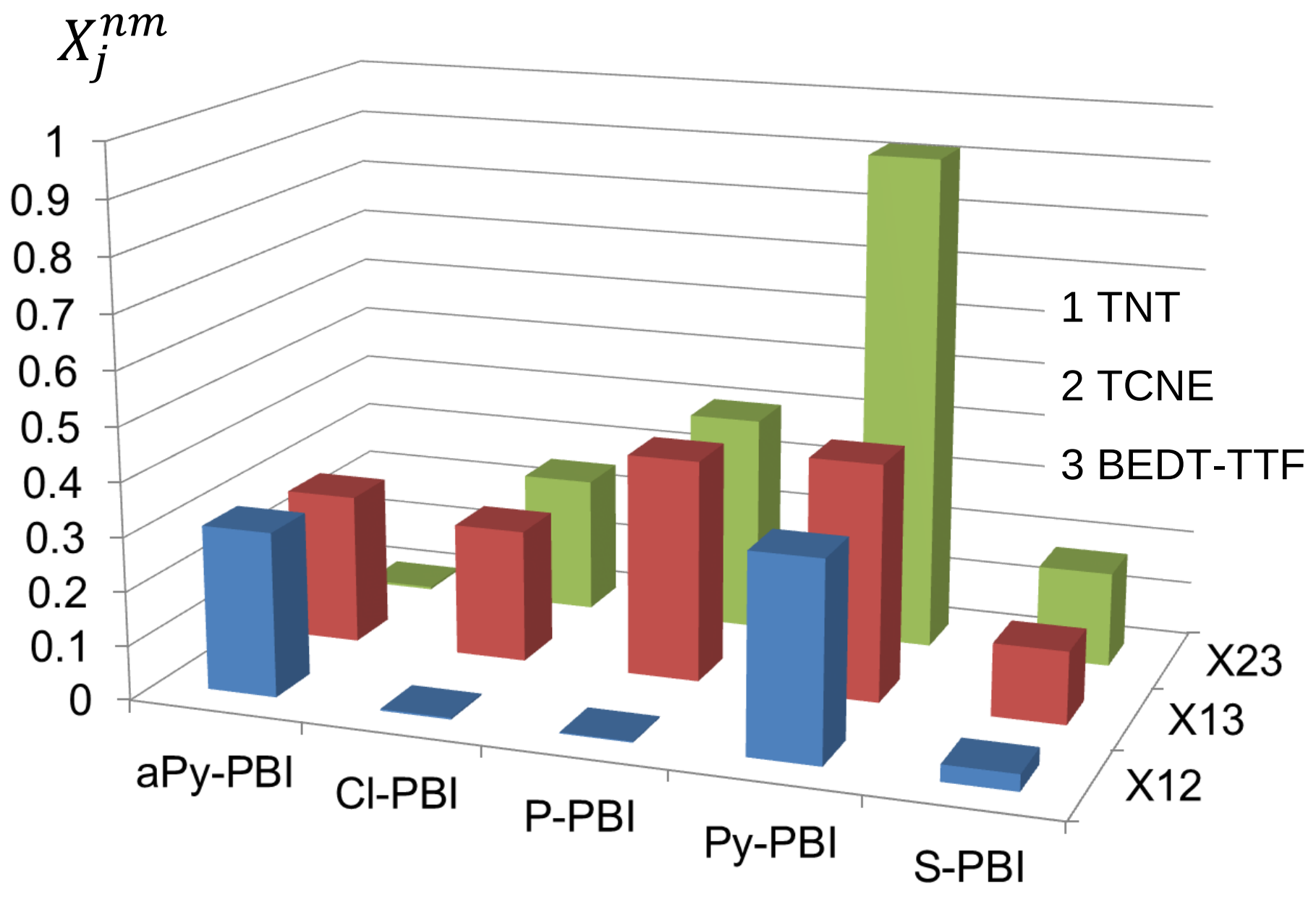

**Figure 14.** Comparison between the normalized correlators $X_j^{nm}$ where $n$ labels the three analytes (TNT, TCNE, and BEDT-TTF) and $j$ labels the five PBIs. V=1.0 volts.

## Movies

The following movies show the different configurations of the analytes on the PBI backbones, used to compute the transmission curves of figure 8 a-c and the I-V curves of figure 8d.

BEDT-TTF-PBI.mpg TCNE-PBI.mpg TNT-PBI.mpg